\documentclass[10pt]{emulateapj}
\usepackage{amsmath}
\usepackage{bm}
\usepackage{pslatex}
\usepackage{graphicx}
\usepackage{natbib}
\begin{document}

\title{Stochastic Heating, Differential Flow, and the Alpha-to-Proton
  Temperature Ratio in the Solar Wind}

\author{B. D. G. Chandran\altaffilmark{1,4},
  D. Verscharen\altaffilmark{1}, E. Quataert\altaffilmark{2},
  J. C. Kasper\altaffilmark{3}, P. A. Isenberg\altaffilmark{1},
\&  S. Bourouaine\altaffilmark{1}}

\altaffiltext{1}{Space Science Center and Department of Physics,  University of New Hampshire, Durham, NH 03824;  benjamin.chandran@unh.edu, s.bourouaine@unh.edu, phil.isenberg@unh.edu, daniel.verscharen@unh.edu}

\altaffiltext{2}{Astronomy Department \& Theoretical Astrophysics  Center, 601 Campbell Hall, The University of California, Berkeley,  CA 94720; eliot@astro.berkeley.edu} 

\altaffiltext{3}{Harvard-Smithsonian Center for Astrophysics, Cambridge, MA 02138; jkasper@cfa.harvard.edu}

\altaffiltext{4}{Merton College, University of Oxford, Oxford OX1 4JD, United Kingdom}

\begin{abstract}
  We extend previous theories of stochastic ion heating to account for
  the motion of ions along the magnetic field~$\bm{B}$. We derive an
  analytic expression for the temperature ratio~$T_{\perp \rm
    i}/T_{\perp \rm p}$ in the solar wind assuming that stochastic
  heating is the dominant ion heating mechanism, where~$T_{\perp \rm
    i}$ is the perpendicular temperature of species~i and $T_{\perp
    \rm p}$ is the perpendicular proton temperature. This expression
  describes how~$T_{\perp \rm i}/T_{\perp \rm p}$ depends upon~$U_{\rm
    i}$ and~$\beta_{\parallel \rm p}$, where~$U_{\rm i}$ is the
  average velocity along~$\bm{B}$ of species~$\mbox{i}$ in the proton
  frame and $\beta_{\parallel \rm p}$ is the ratio of the parallel
  proton pressure to the magnetic pressure, which we take to
  be~$\lesssim 1$.  We compare our model with previously published
  measurements of alpha particles and protons from the~{\em Wind}
  spacecraft. We find that stochastic heating offers a promising
  explanation for the dependence of $T_{\perp \alpha}/T_{\perp \rm p}$
  on $U_\alpha$ and $\beta_{\parallel \rm p}$ when the fractional
  cross helicity and Alfv\'en ratio at the proton-gyroradius scale
  have values that are broadly consistent with solar-wind
  measurements. We also predict how the temperatures of other ion
  species depend on their drift~speeds.
\end{abstract} \keywords{solar wind --- Sun: corona --- turbulence --- waves}

\maketitle

\vspace{0.2cm} 
\section{Introduction}
\label{sec:intro}
\vspace{0.2cm} 

As solar-wind plasma flows away from the Sun, it moves into regions of
progressively weaker magnetic field. If the magnetic moments~$\mu$ of
solar-wind ions were conserved, then the perpendicular
temperature~$T_\perp$ of each ion species would be a strongly decreasing
function of heliocentric distance~$r$. However, in situ spacecraft
measurements show that~$T_\perp$ decreases much more slowly with~$r$
than $\mu$ conservation would imply, indicating that ions undergo some
form of perpendicular heating~\citep{marsch82a,marsch82b,kohl98}.

One model for explaining this heating invokes resonant cyclotron
heating by Alfv\'en/ion-cyclotron (A/IC)
waves~\citep{hollweg02}. Cyclotron heating violates $\mu$ conservation
and offers a possible explanation for the observed preferential
heating of minor ions~\citep{isenberg07} as well as the shape of the
core of the proton velocity distribution in fast-solar-wind
streams~\citep{galinsky00,isenberg01,marsch04,isenberg11}.  The
primary difficulty faced by cyclotron-heating models is that it is not
clear that high-frequency A/IC waves can be produced in sufficient
quantities to explain the levels of proton heating that are observed.
Early studies postulated that a turbulent cascade efficiently
transfers the energy of non-compressive, magnetohydrodynamic (MHD)
fluctuations from large scales to small scales and from low
frequencies to high
frequencies~\citep{isenberg83,isenberg84,tu84,hollweg86,hollweg88}. Since
there is abundant energy in non-compressive, large-scale, MHD-like
fluctuations in the solar wind~\citep{belcher71}, such a cascade would
lead to substantial energy in high-frequency A/IC waves. Theoretical
and numerical investigations, however, have shown that the energy of
non-compressive MHD fluctuations cascades primarily to
smaller~$\lambda_\perp$ and only weakly to
smaller~$\lambda_\parallel$, where $\lambda_\perp$ and
$\lambda_\parallel$ are lengthscales perpendicular and parallel to the
magnetic field~$\bm{B}$~\citep{shebalin83,goldreich95,maron01}. Since
the linear Alfv\'en wave frequency is~$\sim v_{\rm
  A}/\lambda_\parallel$, where $v_{\rm A}$ is the Alfv\'en speed, the
absence of an efficient cascade to small~$\lambda_\parallel$ implies
that the cascade of energy from low frequency to high frequency is
inefficient.

An alternative mechanism for generating high-frequency A/IC waves is
through a turbulent cascade involving compressive magnetosonic
waves. The energy of fast magnetosonic waves (``fast waves'') cascades
radially in wavenumber space, from low frequency to high
frequency~\citep{cho02}.  When the angle between the wavevector and
the background magnetic field is small, high-frequency fast waves
efficiently generate high-frequency Alfv\'en
waves~\citep{chandran05a,chandran08b}. While these mechanisms could
potentially be important in the corona~\citep{lih01,cranmer12}, the
observed anti-correlation between density fluctuations and
magnetic-field-strength fluctuations in the solar wind at 1~AU implies
that fast waves comprise only a tiny fraction of the energy of the
turbulence~\citep{yao11,howes12}. This observational finding makes it
unlikely that a fast-wave cascade can be a significant source of
high-frequency fluctuations in the near-Earth solar wind.  Another
source of high-frequency A/IC waves is plasma
instabilities~\citep{kasper02,hellinger06,matteini07,bale09,wicks10}.
Instabilities driven by proton beams or alpha-particle beams could
thermalize the beam energy and lead to substantial perpendicular ion
heating~\citep{gary00,hellinger11,verscharen13a}.  The extent to which
beam-driven instabilities contribute to perpendicular ion heating in
the solar wind remains an open question.

Dissipation of low-frequency turbulence, rather than high-frequency
A/IC waves, offers an alternative explanation for perpendicular ion
heating in the solar wind. As mentioned previously, most of the
fluctuation energy at small scales in solar-wind turbulence is
believed to consist of non-compressive fluctuations with
$\lambda_\perp \ll
\lambda_\parallel$~\citep{matthaeus90,bieber94,chen12}. If the
dissipation of such fluctuations at small scales proceeded via linear
wave damping, the result would be parallel ion heating and parallel
electron heating rather than perpendicular ion
heating~\citep{quataert98}. On the other hand, a number of studies
have shown that the dissipation of low-frequency turbulence via
nonlinear mechanisms can violate $\mu$
conservation~\citep{dmitruk04,parashar09,servidio12}. One such mechanism is
stochastic
heating~\citep{mcchesney87,karimabadi94,chen01,johnson01,chaston04,fiksel09,bourouaine13}. In
stochastic ion heating, fluctuations at scales comparable to the ion
gyroradii cause ion orbits to become disordered or stochastic in the
plane perpendicular to~$\bm{B}$, violating one of the preconditions
for $\mu$ conservation~\citep{kruskal62}. The interactions of such
ions with the time-varying electric field cause ions to diffuse in
perpendicular kinetic energy, leading to perpendicular heating.

Recently, \cite{kasper13} (hereafter K13) presented detailed observations of the
alpha-to-proton perpendicular temperature ratio~$T_{\perp \alpha}/T_{\perp \rm p}$ 
 in the solar wind, and described how this
temperature ratio
depends upon the average flow velocity~$U_\alpha$ of alpha
particles in the proton frame as well as
\begin{equation}
\beta_{\parallel \rm p} = \frac{8 \pi n_{\rm p} k_{\rm B} T_{\parallel \rm p}}{B_0^2}.
\label{eq:defbetap} 
\end{equation} 
They found that $T_{\perp \alpha}/T_{\perp \rm p} > 6$ when
$U_{\alpha} \ll v_{\rm A}$, that $T_{\perp \alpha}/T_{\perp \rm p}$
decreases to values of~$\sim 4$ as $U_{\alpha}/v_{\rm A}$ increases
towards unity when~$\beta_{\parallel \rm p} \ll 1$, and that the
decrease in $T_{\perp \alpha}/T_{\perp \rm p}$ with
increasing~$U_{\alpha}/v_{\rm A}$ is much less pronounced
when~$\beta_{\parallel \rm p} \gtrsim 1$.  (These general trends are
illustrated in Figure~\ref{fig:Kasper_Talpha3}.)  K13 argued that
these observations are consistent with Isenberg \& Vasquez's (2007)
model \nocite{isenberg07} of resonant cyclotron heating of heavy ions.
In this paper, we address the question of whether these observations
are consistent with stochastic ion heating by low-frequency
turbulence.  \cite{chandran10a} predicted that stochastic ion heating
is less effective when the ions stream away from the Sun in the proton
frame if the majority of the Alfv\'en-wave-like fluctuations propagate
away from the Sun in the proton frame. This is qualitatively
consistent with K13's observations.  In this paper, we develop this
idea in a more quantitative fashion to enable a better comparison
between the stochastic-heating model and spacecraft observations.  In
Section~\ref{sec:theory}, we extend the stochastic-heating theory
developed by \cite{chandran10a} to account for the motion of ions
along~$\bm{B}$.  In Section~\ref{sec:ratios}, we use these results to
derive an analytic expression for the temperature ratios of different
ion species in the solar wind. We then present theoretical
calculations of how the alpha-to-proton temperature ratio depends upon
$U_\alpha$, $\beta_{\parallel \rm p}$, and various properties of the
turbulent fluctuations. We discuss our results and summarize our
conclusions in Section~\ref{sec:conclusion}.

\vspace{0.2cm} 
\section{Stochastic Heating and Parallel Ion Motion}
\label{sec:theory} 
\vspace{0.2cm}

We model the solar wind as a magnetized plasma containing
low-frequency, quasi-2D turbulence.
At perpendicular (parallel) lengthscales~$\lambda_{\perp}$ ($\lambda_\parallel$)
satisfying $\rho_{\rm p} \ll \lambda_\perp \ll L$, this turbulence has the
properties of reduced magnetohydrodynamic (RMHD) turbulence, including
the inequality~$\lambda_\perp \ll \lambda_\parallel$, where
$\rho_{\rm p}$ is the rms gyroradius of the full proton velocity
distribution, and $L$ is the outer scale of the turbulence.
 We assume that the turbulence
transitions to kinetic Alfv\'en wave (KAW) turbulence at
lengthscales~$\lesssim \rho_{\rm p}$. By invoking the term ``kinetic
Alfv\'en wave,'' we do not mean to imply that the fluctuations oscillate
monochromatically or that the turbulence is weak.
Instead, we assume that the turbulence is strong and
``critically balanced,'' in the sense that the linear and nonlinear
operators in the governing equations are of comparable
importance~\citep{higdon84,goldreich95,cho04b,horbury08,schekochihin09}.  The
label ``KAW'' in the term ``KAW turbulence'' refers to the nature of
these linear operators, whose eigenfunctions correspond to KAWs.

We now consider the stochastic heating of ions with mass~$Am_{\rm p}$
and charge~$Ze$, where $m_{\rm p}$ and~$e$ are the proton mass and
charge.  We make no restriction on whether these ions are minor ions,
alpha particles, or protons.  For the moment, we focus on ions with
velocity component~$v_\parallel$ parallel to the background magnetic
field~$\bm{B}_0$ and take $v_\parallel$ to be approximately constant
over the stochastic-heating timescale, where~$v_\parallel$ is measured
in the average proton rest frame. Eventually, we will average
over~$v_\parallel$, but for now we define $Q_\perp(v_\parallel)$,
$T_\perp(v_\parallel)$, and $\rho(v_\parallel)$ to be the stochastic
heating rate, perpendicular temperature, and rms gyroradius of ions
with parallel velocity~$v_\parallel$, where $\rho(v_\parallel) =
v_\perp(v_\parallel)/\Omega_{\rm i}$,
\begin{equation}
v_\perp(v_\parallel) = \sqrt{\frac{2 k_{\rm
    B} T_\perp(v_\parallel)}{Am_{\rm p}}},
\label{eq:defvperp} 
\end{equation}  
and $\Omega_{\rm i} = ZeB_0/(Am_{\rm p} c)$ is the ion cyclotron frequency.

Although the ions may interact with turbulent fluctuations over a
broad range of lengthscales, we only consider the contribution
to~$Q_\perp(v_\parallel)$ from the electric and magnetic-field
fluctuations at lengthscales~$\sim\rho(v_\parallel)$.  We define
$\delta \bm{E}$ and $\delta \bm{B}$ to be the gyroscale electric and
magnetic fields in the average proton rest frame.  The quantity
\begin{equation}
\delta \bm{v} = c\,\frac{\delta \bm{E}\times\bm{B}}{B^2}
\label{eq:defdv} 
\end{equation} 
is then the gyroscale $\bm{E}\times\bm{B}$ velocity in the average
proton frame. We assume that $\delta B \ll B_0$.
Neglecting the component of~$\delta \bm{E}$ parallel to the magnetic
field and corrections of order~$\delta B/B_0$, we rewrite
Equation~(\ref{eq:defdv}) in the form
$\delta \bm{E} = -\, \delta \bm{v}\times \bm{B}_0/c$.
We define the Els\"asser variables in the average proton rest frame
as
\begin{equation}
\bm{z}^\pm = \delta \bm{v} \mp v_{\rm A} \frac{\delta \bm{B}_\perp}{B_0},
\label{eq:defzpm} 
\end{equation} 
where $\delta\bm{B}_\perp = \delta \bm{B} -
\bm{\hat{b}}(\bm{\hat{b}}\cdot \delta \bm{B})$, $\bm{\hat{b}} =
\bm{B}_0/B_0$, $v_{\rm A} = B_0/\sqrt{4\pi n_{\rm p} m_{\rm p}}$ is the
(proton) Alfv\'en speed, and $n_{\rm p}$ is the proton number density.

We define the ``$v_\parallel$ frame'' to be the reference frame moving
at velocity~$v_\parallel \bm{\hat{b}}_0$ with respect to the average
proton rest frame.  The gyroscale electric-field fluctuation in this
frame is given by $\delta \bm{E}^\prime = \delta \bm{E} + v_\parallel
\bm{\hat{b}} \times \delta \bm{B}/c$. We take $|v_\parallel|$
and~$v_{\rm A}$ to be~$\ll c$ and thus neglect the difference
between~$\bm{B}$ in the $v_\parallel$ frame 
and~$\bm{B}$ in the proton frame.  The
$\bm{E}\times\bm{B}$ velocity in the $v_\parallel$~frame is then given
by $\delta \bm{v}^\prime = c \delta \bm{E}^\prime \times
\bm{B}/B^2$. Neglecting corrections of order $\delta B/B_0$, we obtain
\begin{equation}
\delta \bm{v}^\prime(v_\parallel) = \frac{\bm{z}^+}{2}\left(1 -
\frac{v_\parallel}{v_{\rm A}}\right) + \frac{\bm{z}^-}{2}\left(1 +
\frac{v_\parallel}{v_{\rm A}}\right).
\label{eq:defvprime} 
\end{equation} 

The stochastic heating rate per unit mass $Q_\perp(v_\parallel)$ can
now be obtained in exactly the same way as in the phenomenological
treatment of \cite{chandran10a}, but replacing their~$\delta v_{\rm
  i}$ (the rms value of $\delta \bm{v}$) with $\delta v^\prime_{\rm
  rms}$, the rms value of $\delta \bm{v}^\prime$. This leads to the
expression
\begin{equation}
Q_\perp(v_\parallel) = \frac{c_1 [\delta v^\prime_{\rm rms}(v_\parallel)]^3}{\rho(v_\parallel)} \,\exp\left(-\,\frac{c_2}{\epsilon}\right),
\label{eq:Qperp1} 
\end{equation} 
where 
\begin{equation}
\epsilon = \frac{\delta v^\prime_{\rm rms}(v_\parallel)}{v_{\perp}(v_\parallel)}.
\label{eq:defeps} 
\end{equation} 
The quantities~$c_1$ and $c_2$ are dimensionless constants of order
unity. The derivation of Equation~(\ref{eq:Qperp1}) assumes that the
ion thermal speeds are~$\lesssim v_{\rm A}$.  Assuming that the ion
temperature anisotropies are not very large and that the ion thermal
speed is comparable to the proton thermal speed, this condition is
approximately equivalent to
\begin{equation}
\beta_{\parallel \rm p} \lesssim 1.
\label{eq:betacond}
\end{equation} 
We return to this condition in Section~\ref{sec:ratios}.

We define the gyroscale fractional cross helicity
\begin{equation}
\sigma = \frac{ \left \langle |\bm{z}^+|^2 - |\bm{z}^-|^2\right\rangle}{ \left \langle |\bm{z}^+|^2 + |\bm{z}^-|^2\right\rangle},
\label{eq:defsigma} 
\end{equation} 
the gyroscale Alfv\'en ratio
\begin{equation}
r_{\rm A} = \left(\frac{B_0}{v_{\rm A}}\right)^2 \frac{\langle |\delta \bm{v}|^2 \rangle}{\langle |\delta \bm{B}_\perp|^2\rangle},
\label{eq:defrA} 
\end{equation} 
and the quantity
\begin{equation}
W = \frac{1}{4} \left\langle |\bm{z}^+|^2 + |\bm{z}^-|^2\right\rangle,
\label{eq:defEturb} 
\end{equation} 
where $\langle \dots \rangle$ indicates a time or volume average.  If
$\rho(v_\parallel) \gg \rho_{\rm p}$, then the gyroscale fluctuations
are in the RMHD regime, $\delta \bm{B}_\perp = \delta \bm{B}$ to a
good approximation, and the $\bm{E}\times\bm{B}$ velocity is
approximately equal to the component of the average proton velocity
perpendicular to~$\bm{B}$. In this case, $W$ is the energy of the
gyroscale fluctuations per unit mass.  On the other hand, if
$\rho(v_\parallel) \simeq \rho_{\rm p}$, then the fluctuations are at
the transition to the KAW regime, $\bm{\hat{b}}\cdot \delta \bm{B}
\sim \delta B_\perp$, the protons do not move at
the $\bm{E}\times \bm{B}$ velocity, and $W$ 
differs (by a factor of order unity) from  the energy per
unit mass~\citep{hollweg99c}.  Upon taking the rms of the right-hand side of
Equation~(\ref{eq:defvprime}), we obtain
\begin{equation}
\delta v^\prime_{\rm rms}(v_\parallel) = \left(\chi W\right)^{1/2},
\label{eq:uprime2} 
\end{equation} 
where
\begin{equation} 
\chi = \frac{2}{r_{\rm A} + 1}\left(r_{\rm A} + \frac{v_\parallel^2}{v_{\rm A}^2}\right)
- \frac{2\sigma v_\parallel}{v_{\rm A}}.
\label{eq:defchi} 
\end{equation}

\vspace{0.2cm} 
\section{Ion Temperature Ratios in the Solar Wind}
\label{sec:ratios} 
\vspace{0.2cm}

\begin{figure}[t]
\centerline{
\includegraphics[width=7cm]{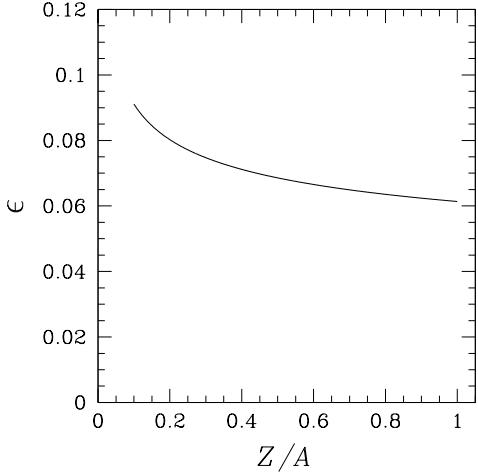}
}
\caption{Stochasticity parameter $\epsilon = \delta v^\prime_{\rm
    rms}(v_\parallel)/v_{\perp}(v_\parallel)$ as a function of $Z/A$,
  which is the ion charge-to-mass ratio in units of the proton
  charge-to-mass ratio. This value of $\epsilon$ is obtained by
  solving the ion energy balance equation, Equation~(\ref{eq:tsc}),
  for the case in which $c_1 =0.74$, $c_2 = 0.21$, $B = 5 \mbox{ nT}$,
  $U+v_\parallel = 400 \mbox{ km/s}$, and $r = 1 \mbox{ AU}$. The weak
  variation of $\epsilon$ with~$Z/A$ justifies our approximation that
  $\epsilon$ is the same for all ions at the same location.
\label{fig:Kasper_eps} }
\end{figure}

In the solar wind at heliocentric distances~$r$ between $0.3$~AU and
1~AU, the perpendicular temperatures of protons and alpha particles
decrease with increasing~$r$, but not as fast as they would decrease
in the case of adiabatic expansion~\citep{marsch82a,marsch82b,hellinger13}. This
implies that there is a competition between adiabatic cooling and
heating and that the heating timescale~$t_{\rm h}$ is of the same
order of magnitude as the cooling or expansion timescale~$t_{\rm exp}
= r/(U+v_\parallel)$, where~$U$ is the proton outflow velocity. We
assume that this is true not just for protons and alpha particles, but
for all ion species. We further assume that the dominant perpendicular
ion heating mechanism is stochastic heating, so that $t_{\rm h} \simeq
v_{\perp i}^2/Q_\perp$, where we have suppressed the dependence of
these quantities on~$v_\parallel$ for brevity. Approximately equating
$t_{\rm h}$ and $t_{\rm exp}$ leads to the condition
\begin{equation}
\epsilon^{-3} \exp\left(\frac{c_2}{\epsilon}\right)
\simeq \frac{c_1 r \Omega_{\rm i}}{U+v_\parallel}.
\label{eq:tsc} 
\end{equation} 
A version of this relation with $c_1 \simeq 1$ and
$v_\parallel = 0$ was previously obtained by \cite{chandran10b} in a
study of ion temperatures in coronal holes.  Because the left-hand
side of Equation~(\ref{eq:tsc}) is a rapidly varying function
of~$\epsilon$, Equation~(\ref{eq:tsc}) leads to similar values
of~$\epsilon$ for different ion species. We illustrate this point in
Figure~\ref{fig:Kasper_eps}, which plots the solution of
Equation~(\ref{eq:tsc}) for plasma parameters characteristic of the
slow solar wind near Earth using the values $c_1 = 0.74$ and $c_2 =
0.21$ obtained in a recent numerical simulation of the stochastic
heating of test-particles in RMHD turbulence~\citep{xia13}.
When Equation~(\ref{eq:tsc}) is
applied to ions of the same species in the solar wind with different
values of~$v_\parallel$, these variations in $v_\parallel$ lead to
fractional variations in the right-hand side of
Equation~(\ref{eq:tsc}) that are small, because the ion thermal speeds
are $\ll U$. Figure~\ref{fig:Kasper_eps} shows that small fractional
variations in the right-hand side of Equation~(\ref{eq:tsc}) lead to
extremely small variations in~$\epsilon$. In the analysis to follow,
we make the approximation that $\epsilon$ is independent
of both~$v_\parallel$ and~$Z/A$.

To illustrate how ions can achieve the value of~$\epsilon$ needed to
satisfy Equation~(\ref{eq:tsc}), we consider the hypothetical
evolution of ions that are initially sufficiently cool that $\epsilon$
is larger than the value in Equation~(\ref{eq:tsc}). For such ions,
$t_{\rm h}$ is initially smaller than~$t_{\rm exp}$, and thus heating
initially dominates over cooling, causing $T_{\perp \rm i}$ to
increase. In the case of minor ions, this heating draws a negligible
amount of power from the turbulence, and $T_\perp(v_\parallel)$ simply
increases until $\epsilon$ decreases to the value in
Equation~(\ref{eq:tsc}).  Stochastic heating cannot
increase~$T_\perp(v_\parallel)$ any further, because a higher
$T_\perp(v_\parallel)$ would imply a smaller value of~$\epsilon$ and
hence an exponentially smaller value of~$Q_\perp(v_\parallel)$.  In
the case of protons and alpha particles, stochastic ion heating can
drain a significant fraction of the turbulent cascade power, thereby
reducing~$\delta v^\prime$.  For these ions, Equation~(\ref{eq:tsc})
is in general satisfied by some combination of heating of the ions and
damping of the gyroscale fluctuations.

We now estimate $T_{\perp \rm i}/T_{\perp \rm p}$, the ion-to-proton
perpendicular temperature ratio for ion species~i, which we take to
have mass~$Am_{\rm p}$ and charge~$Ze$, as above. We do not attempt to predict
the way that proton and alpha-particle heating alter the turbulent
power spectrum at wavenumbers~$\sim \rho_{\rm p}^{-1}$. Instead, we
parametrize the dependence of~$W$ on $\rho(v_\parallel)$ through the
equation
\begin{equation}
W  = W_{\rm p} \left[\frac{\rho(v_\parallel)}{\rho_{\rm p}}\right]^{2a},
\label{eq:Wscal} 
\end{equation} 
where the constant~$a$ is a free parameter, and ~$W_{\rm p}$ is a
constant whose value has no influence on our estimate of~$T_{\perp \rm
  i}/T_{\perp \rm p}$. If $\rho(v_\parallel)$ and $\rho_{\rm p}$ were
both in the inertial range of solar-wind turbulence, and if the total
energy spectrum were $\propto k_\perp^{-3/2}$ in the inertial range,
then $a$ would be~$1/4$.  However, we assume that $\rho(v_\parallel)$
is in the range of $\rho_{\rm p}$ to a few~$\rho_{\rm p}$. At
perpendicular wavenumbers~$\sim \rho_{\rm p}^{-1}$, there are two
effects that influence the value of~$a$. First, dissipation acts to
steepen the power spectra of~$\bm{E}$ and~$\bm{B}$ fluctuations,
thereby acting to increase~$a$. Second, the cross-field motions of
electrons and protons are only partially coupled. This decoupling
causes the electric-field (magnetic-field) power spectrum to flatten
(steepen) at $k_\perp \gtrsim \rho_{\rm
  p}^{-1}$~\citep{bale05,howes08b,schekochihin09}. Because $W$
contains contributions from both $\bm{E} \times \bm{B}$ velocity
fluctuations and $\bm{B}$ fluctuations, we conjecture that the
decoupling of electrons and protons at scales~$\sim \rho_{\rm p}$ acts
to reduce~$a$ slightly relative to the value that would apply at
inertial-range scales. We do not attempt to take these various effects
into account and determine the most accurate or physically reasonable
value of~$a$. Instead, we consider a few different values as numerical
examples, focusing on the case in which~$a=1/4$.  To further simplify
the analysis, we treat~$r_{\rm A}$ and~$\sigma$ as constants, thereby
neglecting any possible variation in the Alfv\'en ratio and fractional
cross helicity over scales between~$\sim \rho_{\rm p}$
and~$\rho(v_\parallel)$.

Using Equation~(\ref{eq:uprime2}), we re-write
Equation~(\ref{eq:defeps}) in the form
\begin{equation}
[v_{\perp \rm i}(v_\parallel)]^2 = \epsilon^{-2\eta} W_{\rm p}^\eta \left(\frac{A}{Zw_{\perp \rm p}}\right)^{2a\eta} \chi^\eta  ,
\label{eq:Tscaling} 
\end{equation} 
where $w_{\perp \rm p}$ is the perpendicular thermal speed of the full proton distribution, and
\begin{equation}
\eta = \frac{1}{1-a}.
\label{eq:defeta} 
\end{equation} 
Upon averaging Equation~(\ref{eq:Tscaling}) over~$v_\parallel$ and making
the approximation that~$\epsilon$ is the same for all values
of~$v_\parallel$, we obtain
\begin{equation}
w_{\perp \rm i}^2 =
\epsilon^{-2\eta} W_{\rm p}^\eta \left(\frac{A}{Zw_{\perp \rm p}}\right)^{2a\eta}\langle \chi^\eta \rangle_{\rm i},
\label{eq:Tscaling2} 
\end{equation} 
where $\langle \dots \rangle_{\rm i}$ indicates an average over the
$v_\parallel$ distribution of species~i, $w_{\perp \rm i}^2 =
\langle [v_{\perp \rm i}(v_\parallel)]^2\rangle_{\rm i}$, and the subscripts
$\mbox{i}=\mbox{p}$ and $\mbox{i}=\alpha$ correspond to protons and alpha particles, respectively.  Applying
Equation~(\ref{eq:Tscaling2}) to the protons, we obtain
\begin{equation}
w_{\perp \rm p}^2 = \epsilon^{-2} W_{\rm p} \langle \chi^\eta\rangle^{1/\eta}_{\rm p}.
\label{eq:Tratiop} 
\end{equation} 
Using  Equation~(\ref{eq:Tratiop}) to eliminate~$W_{\rm p}$ in Equation~(\ref{eq:Tscaling2}),
we find that
\begin{equation}
\frac{T_{\perp \rm i}}{T_{\perp \rm p}} = A \left(\frac{A}{Z}\right)^{2a\eta} 
\frac{\langle \chi^\eta\rangle_{\rm i}}{\langle \chi^\eta\rangle_{\rm p}},
\label{eq:Tratio} 
\end{equation} 
where $T_{\perp \rm i}$ and $T_{\perp \rm p}$ (without the functional
dependence on~$v_\parallel$) are the perpendicular temperatures of the
full ion and proton distributions.  We take the $v_\parallel$
distribution of each ion species to be a shifted Maxwellian with
average parallel velocity~$U_{\rm i}$ (where $U_{\rm p}=0$), parallel
temperature~$T_{\parallel \rm i}$, and parallel thermal speed
defined by $w_{\parallel \rm i} = \sqrt{2 k_{\rm B} T_{\parallel \rm
    i}/(A m_{\rm p})}$. Thus,
\begin{equation}
\langle \chi^{\eta} \rangle_{\rm i} = \frac{1}{\sqrt{\pi}\, w_{\parallel\rm i}}
\int_{-\infty}^{\infty} dv_\parallel\,\, \chi^{\eta}\exp\left(-\frac{(v_\parallel - U_{\rm i})^2}{w_{\parallel\rm i}^2}\right).
\label{eq:chiav_alpha} 
\end{equation}

\begin{figure*}[t]
\centerline{
\includegraphics[width=6.5cm]{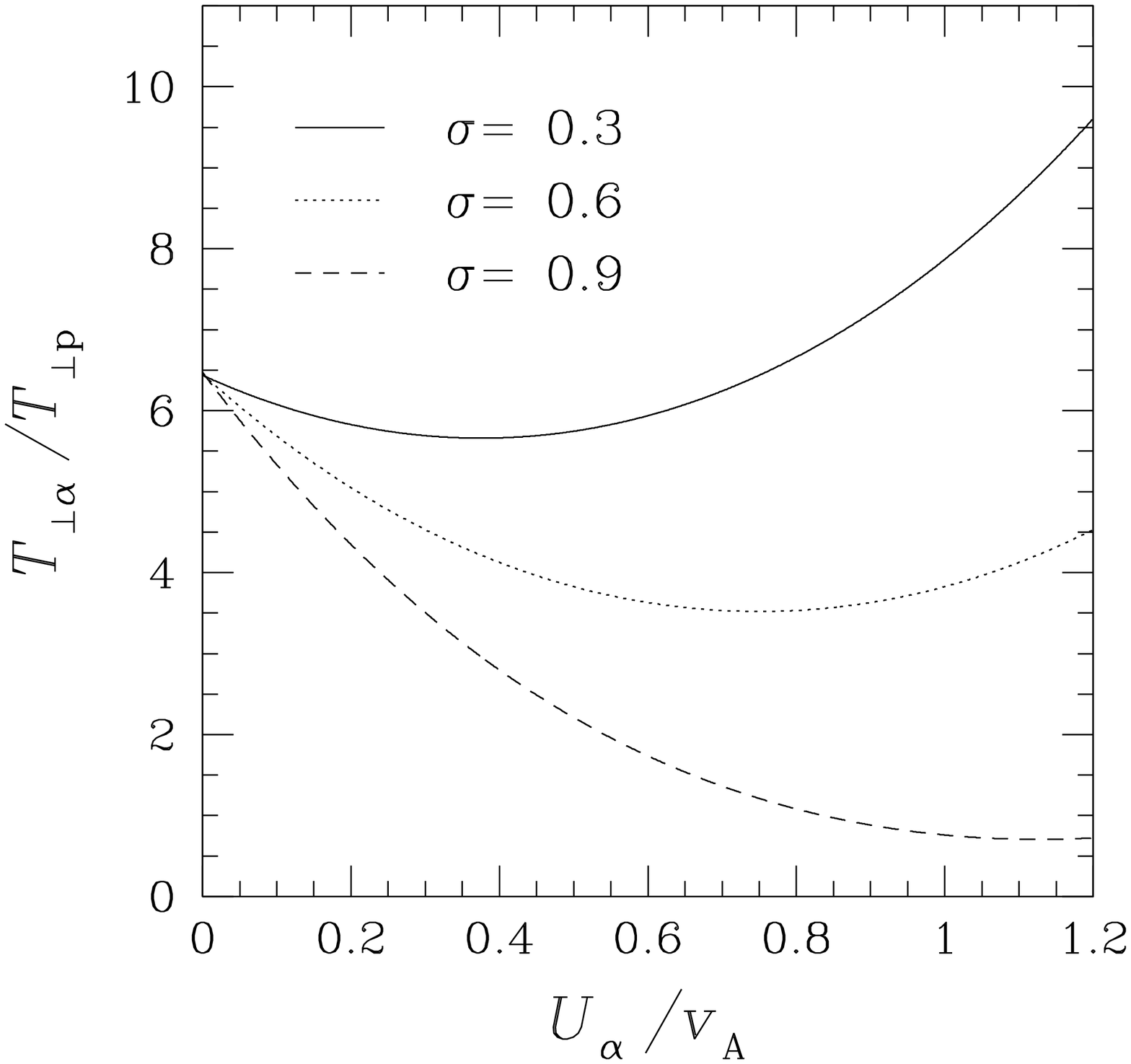}
\hspace{0.5cm} 
\includegraphics[width=6.5cm]{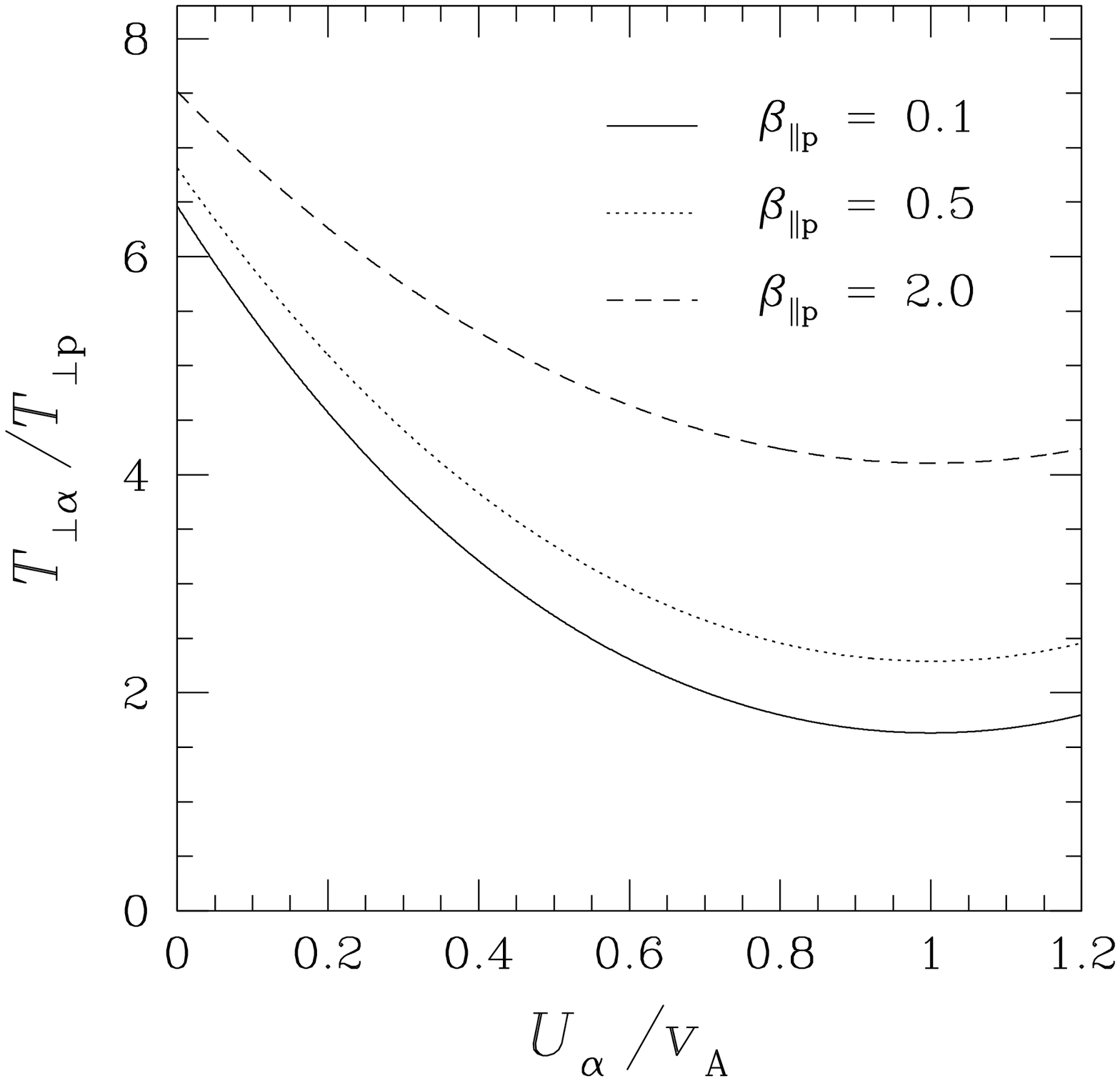} 
}
\centerline{
\includegraphics[width=6.5cm]{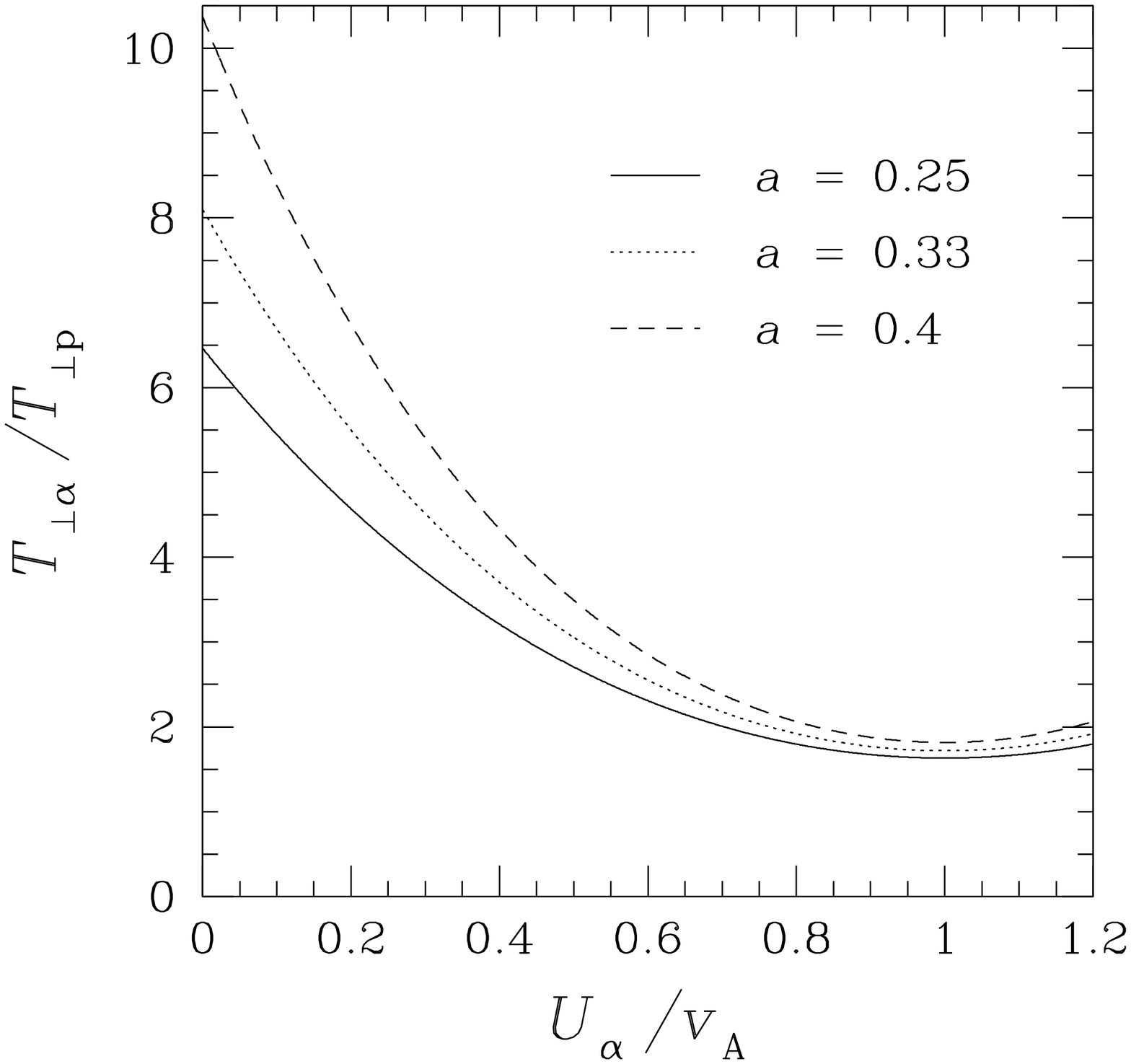}
\hspace{0.5cm} 
\includegraphics[width=6.5cm]{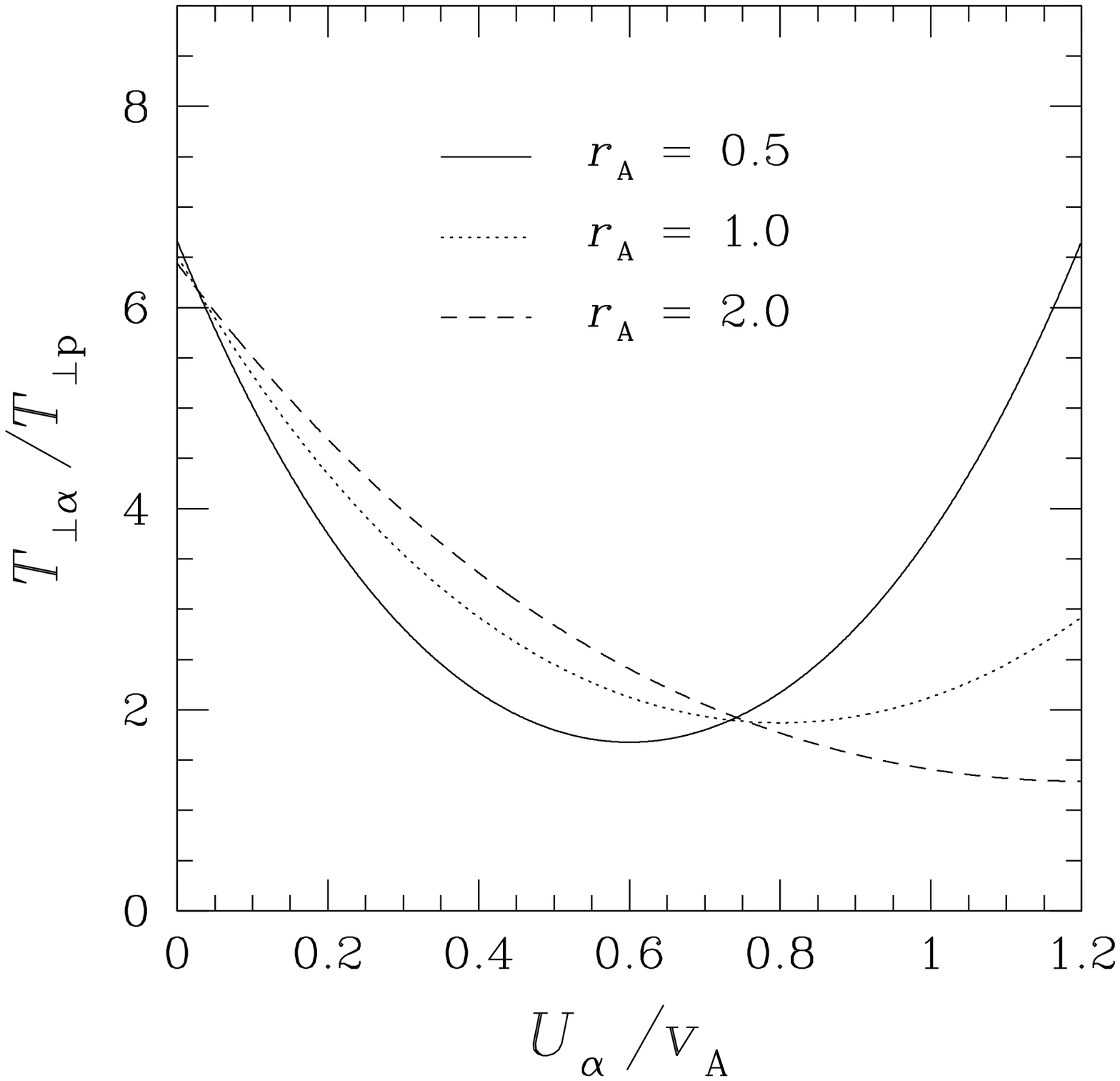} 
}
\caption{ Dependence of the alpha-to-proton perpendicular temperature
  ratio on the alpha-proton drift velocity~$U_\alpha$, the gyroscale
  fractional cross helicity~$\sigma$, the value of $\beta_{\parallel
    \rm p}$, the scaling exponent~$a$ of the rms fluctuation amplitude
  at scales~$\sim \rho_{\rm p}$, and the gyroscale Alfv\'en
  ratio~$r_{\rm A}$. For each plot, $T_{\parallel \alpha} =
  5.2T_{\parallel \rm p}$. Except where indicated otherwise, these
  figures make use of the fiducial parameter values $\sigma = 0.8$,
  $\beta_{\parallel \rm p} = 0.1$, $a = 0.25$, and~$r_{\rm A} = 1.5$.
\label{fig:functional_dependencies} }
\end{figure*}

We now use Equation~(\ref{eq:Tratio}) to calculate the alpha-to-proton
perpendicular temperature ratio. In Figure~\ref{fig:functional_dependencies}, we
illustrate how $T_{\perp \alpha}/T_{\perp \rm p}$ in our model depends upon
$U_\alpha$, $\sigma$, $r_{\rm A}$, $a$, and $\beta_{\parallel \rm p}$ for the
case in which
\begin{equation}
T_{\parallel \alpha} = 5.2 T_{\parallel \rm p},
\label{eq:Tparratio} 
\end{equation} 
which corresponds to the average parallel temperature ratio measured by the {\em
  Wind} spacecraft in the weakly collisional solar-wind streams
examined by~\cite{kasper08}.  As shown in the upper-left panel of
Figure~\ref{fig:functional_dependencies}, if $\sigma$ is close to
one, $\beta_{\parallel \rm p} = 0.1$, and $r_{\rm A} = 1.5$, then
$T_{\perp \alpha}/T_{\perp \rm p}$ undergoes a marked decrease as
$U_\alpha/v_{\rm A}$ increases from zero to~one, because the
transformation from the average proton frame to the average
alpha-particle frame reduces the amplitude of
the electric field fluctuations, which are the source of the
heating. One way of viewing this is through the expression for the
gyroscale $\bm{E}\times \bm{B}$ velocity $\delta v^\prime$ in the
$v_\parallel$ frame given in Equation~(\ref{eq:defvprime}). As
$v_\parallel/v_{\rm A}$ increases from 0 to~1, the contribution of the
anti-sunward-propagating $z^+$ fluctuations to $\delta v^\prime$
decreases, and the contribution to $\delta v^\prime$ from the
sunward-propagating $z^-$ fluctuations increases. When $\sigma $ is
close to~1, the dominant effect is the reduction in the contribution
from the $z^+$ fluctuations. In contrast, as $\sigma$ decreases to
zero, the increase in the contribution to $\delta v^\prime$ from $z^-$
becomes increasingly important and can even cause $\delta
v^\prime_{\rm rms}$ to increase as $v_\parallel$ increases
towards~$v_{\rm A}$.

The upper-right panel of Figure~\ref{fig:functional_dependencies}
shows how the alpha-to-proton temperature ratio depends
upon~$U_\alpha$ for different values of~$\beta_{\parallel \rm
  p}$. When $\beta_{\parallel \rm p}$ is small, the bulk of the alpha
particle distribution can be taken to have velocities $\simeq
U_{\alpha}$. Thus, as $U_{\alpha}$ increases from 0 to~$v_{\rm A}$,
the electric field fluctuations seen by the alphas weaken considerably
when $\sigma $ is close to one, causing $T_{\perp \alpha}/T_{\perp \rm
  p}$ to decrease. On the other hand, if $\beta_{\parallel \rm p }
\gtrsim 1$, then the parallel thermal speed exceeds~$v_{\rm A}$, the
particles are never localized within a narrow band of
$v_\parallel/v_{\rm A}$ values, and there is thus less of a reduction
in $T_{\perp \alpha}/T_{\perp \rm p}$ as $U_{\alpha}/v_{\rm A}$
increases from~0 to~1.  Also, when~$\beta_{\parallel \rm p} \gtrsim 1$
and $T_{\parallel \alpha} > 4 T_{\parallel \rm p}$, alpha particles
are more likely than protons to satisfy either~$v_\parallel < 0$ or
$v_\parallel > v_{\rm A}$, which enhances~$T_{\perp \alpha}/T_{\perp
  \rm p}$ relative to the low-$\beta_{\parallel \rm p}$ case at all
values of~$U_{\alpha}/v_{\rm A}$.

The lower-left panel of Figure~\ref{fig:functional_dependencies} shows
how the alpha-to-proton temperature ratio depends upon~$U_\alpha$ for
different values of~$a$. Larger values of~$a$ correspond to a steeper
turbulent power spectrum at wavenumbers of order $\rho_{\rm
  p}^{-1}$. Because the alpha-particle gyroradii are larger
than~$\rho_{\rm p}$ in the solar wind, increasing~$a$ increases the
amplitudes of the fluctuations that heat the alpha particles relative
to the amplitudes of the fluctuations that heat the protons, thereby
increasing~$T_{\perp \alpha}/T_{\perp \rm p}$. This effect weakens in
the lower-left panel of Figure~\ref{fig:functional_dependencies} as
$U_{\alpha}/v_{\rm A}\rightarrow 1$ because $T_{\perp
  \alpha}/T_{\perp \rm p}$ decreases, thereby reducing the
alpha-particle gyroradii.

The lower-right panel of Figure~\ref{fig:functional_dependencies}
shows how the alpha-to-proton temperature ratio depends
upon~$U_\alpha$ for different values of~$r_{\rm A}$. The
magnetic-field fluctuation in the average proton rest frame,~$\delta
\bm{B}$, does not contribute to particle energization in
the average proton rest frame, since the time derivative of the
particle kinetic energy is simply $Ze \bm{v} \cdot \delta
\bm{E}$. However, the electric field in the $v_\parallel$ frame is $
\delta \bm{E} + \Delta \bm{E}$, where $ \Delta \bm{E} = v_\parallel
\bm{\hat{b}} \times \delta \bm{B}/c$, and thus $\delta \bm{B}$ can
contribute to the heating rate in the $v_\parallel $ frame through the
action of~$\Delta \bm{E}$.  The gyroscale Alfv\'en
ratio~$r_{\rm A}$ controls the relative contributions of $\Delta
\bm{E}$ and $\delta \bm{E}$ to the stochastic heating rate. As $r_{\rm
  A}$ is decreased, $\Delta \bm{E}$ becomes larger relative to~$\delta
\bm{E}$, which can lead to an increase in the total stochastic heating
rate as $v_\parallel/v_{\rm A}$ is increased to values~$\sim 1$. On
the other hand, as $r_{\rm A}$ is increased, the contribution of
$\Delta \bm{E}$ to particle heating becomes less important, and the
stochastic heating rate undergoes a smaller increase as $v_\parallel$
is increased towards~$v_{\rm A}$.

\begin{figure*}[t]
\centerline{
\includegraphics[width=5.5cm]{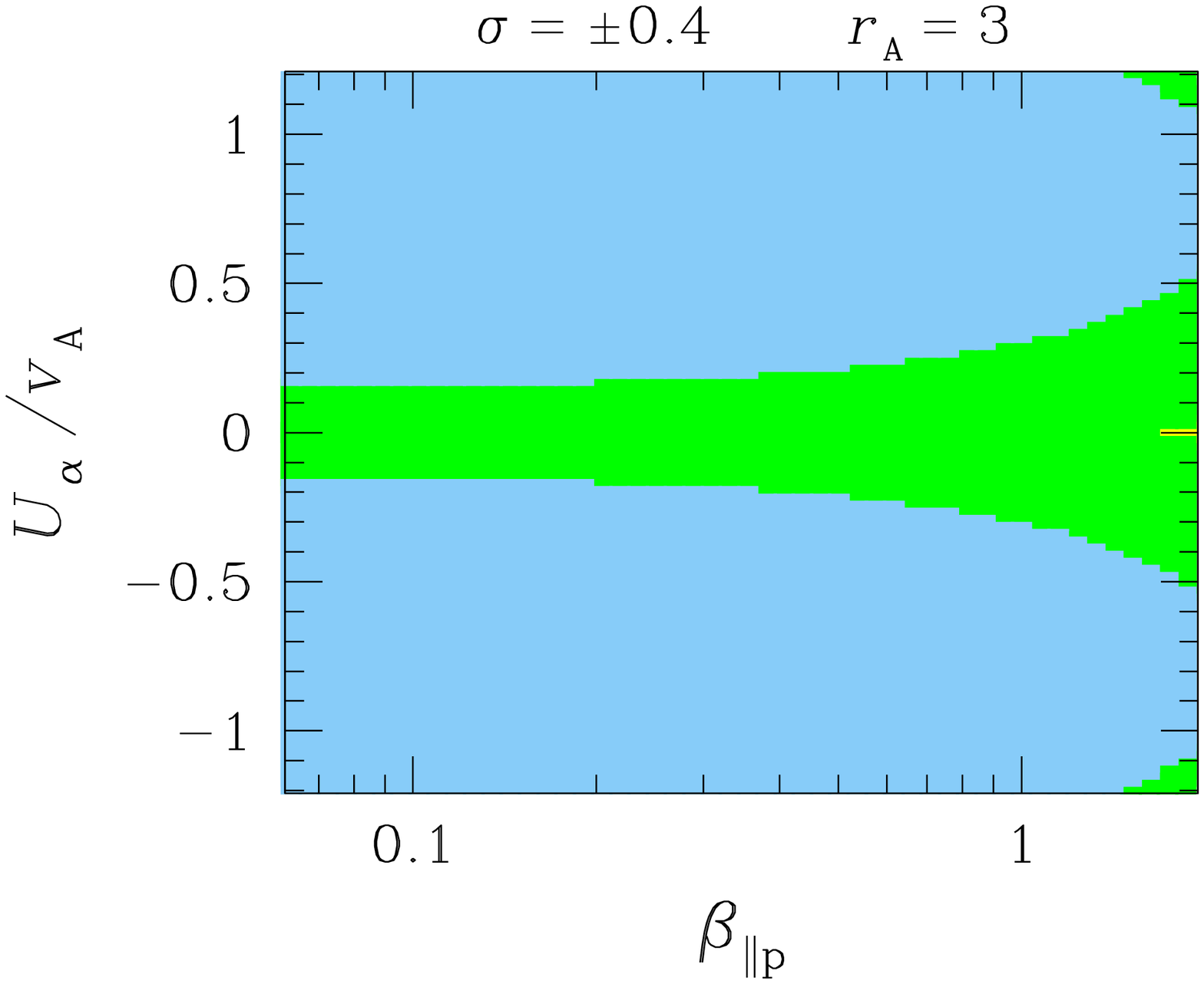}
\includegraphics[width=5.5cm]{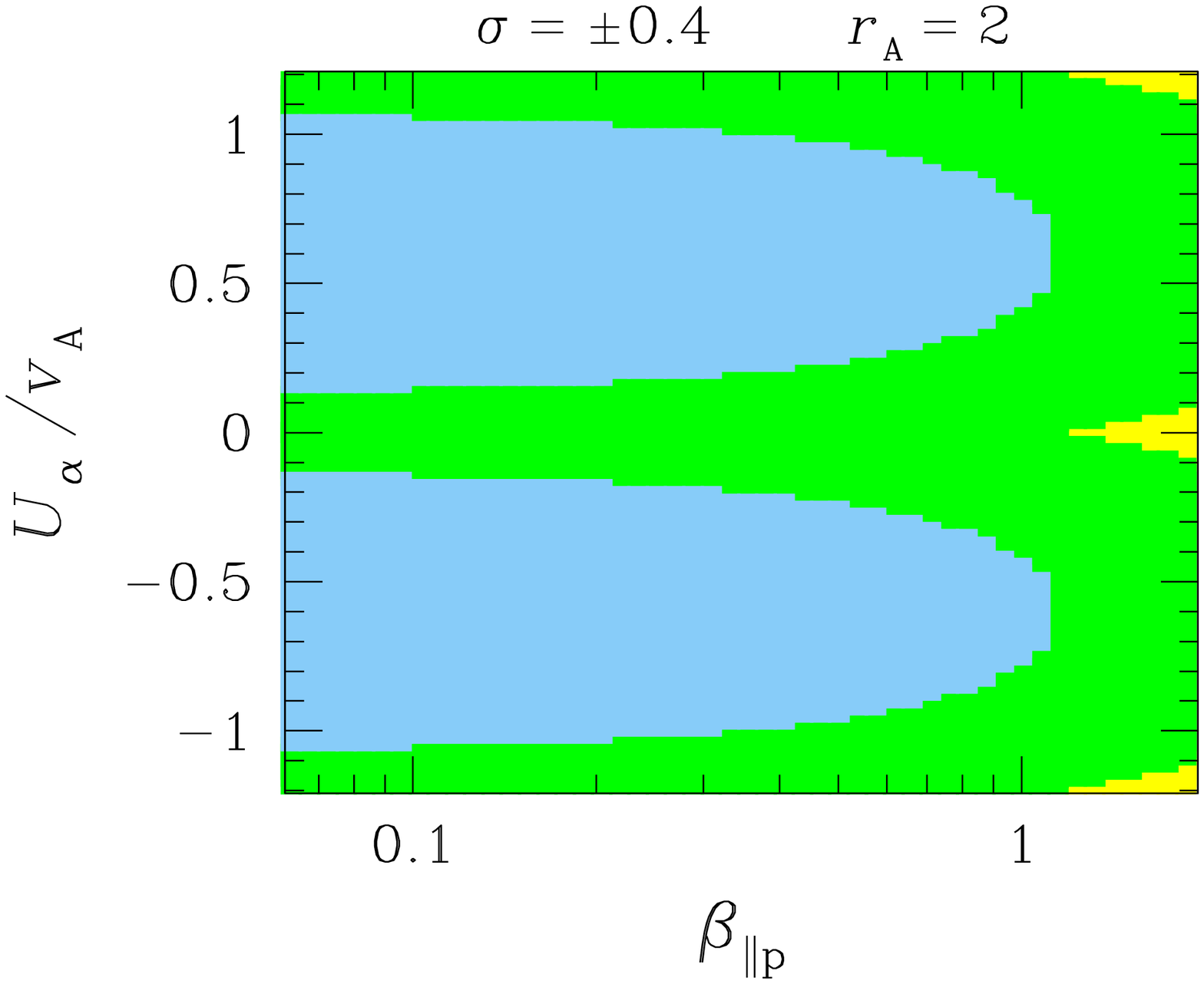}
\includegraphics[width=5.5cm]{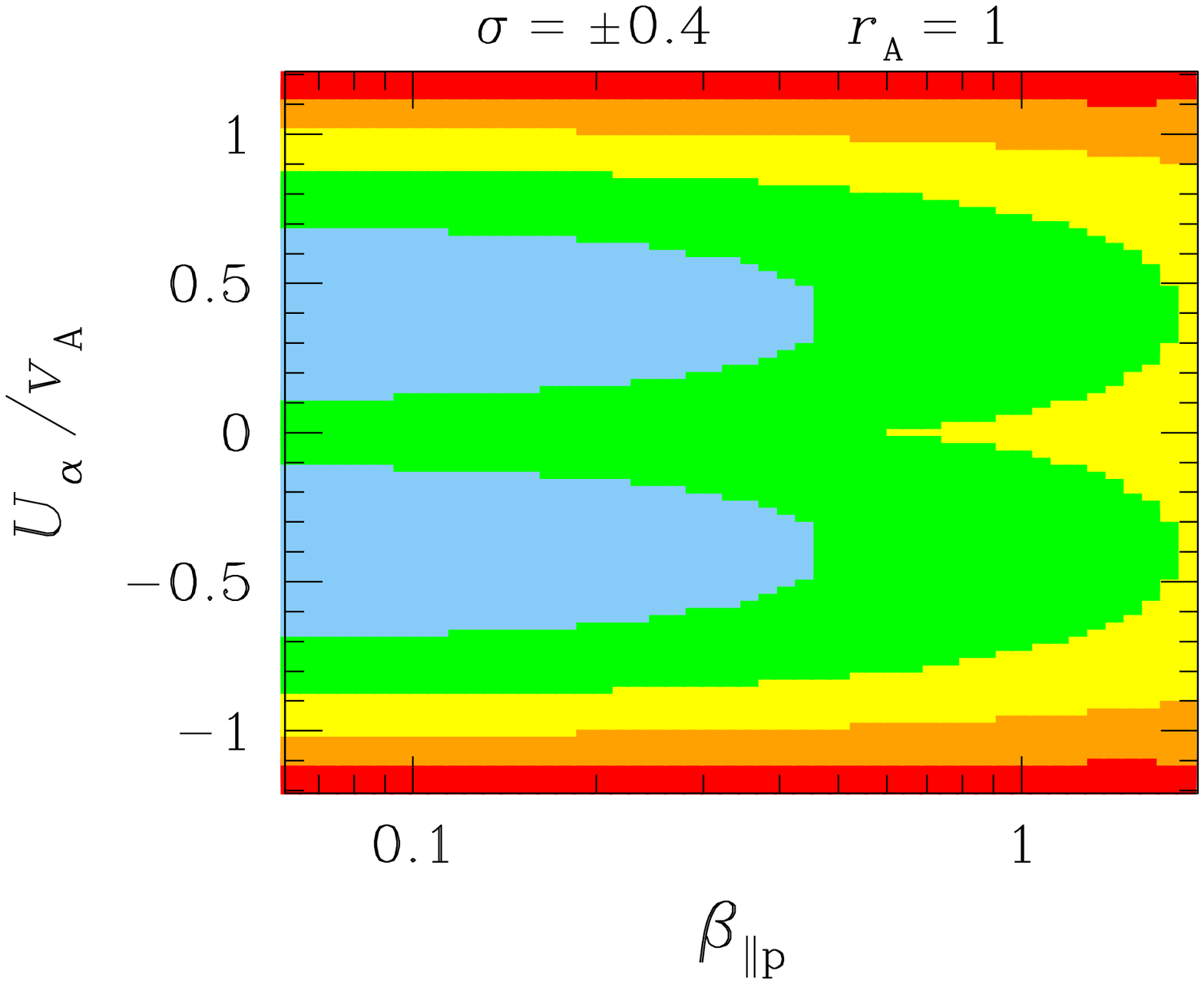}
\hspace{-4.5cm} 
\includegraphics[width=5.5cm]{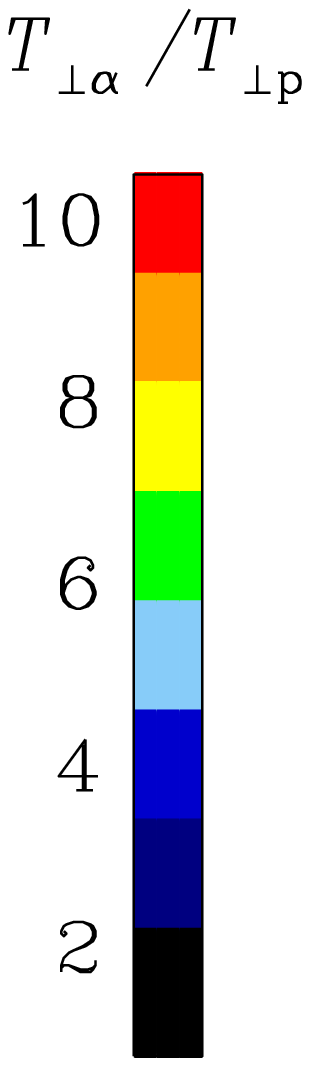}
}
\vspace{-1cm} 
\centerline{
\includegraphics[width=5.5cm]{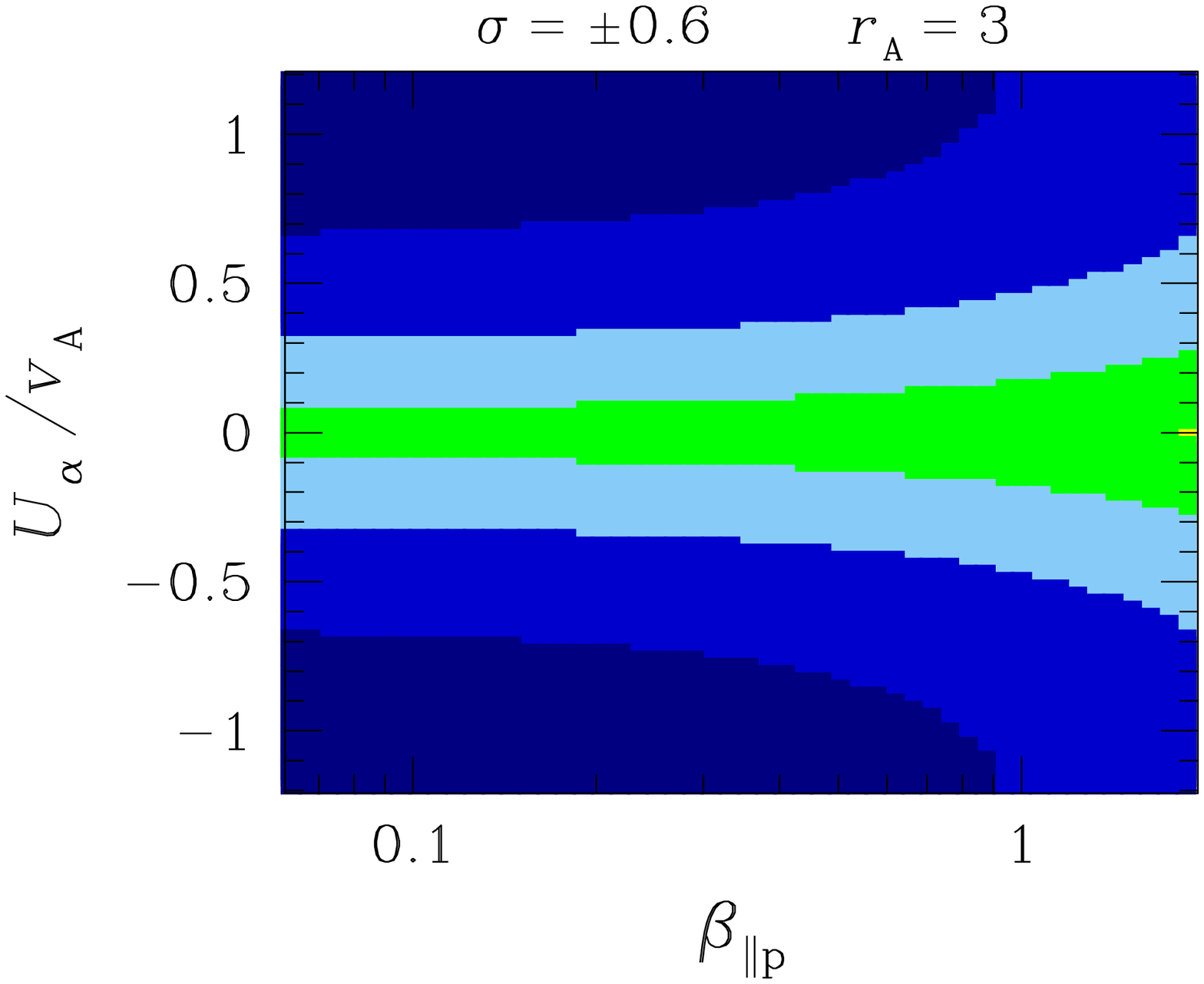}
\includegraphics[width=5.5cm]{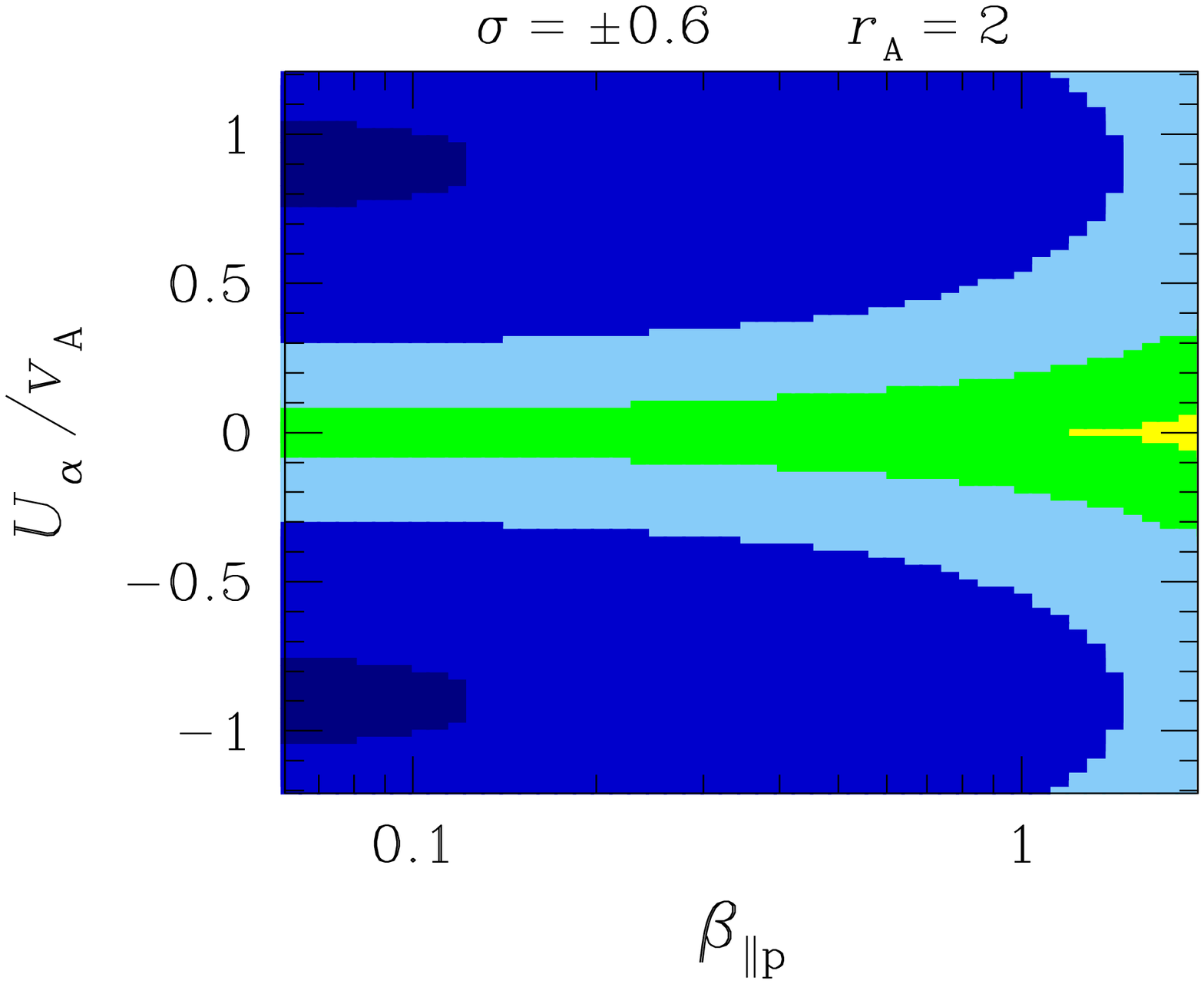}
\includegraphics[width=5.5cm]{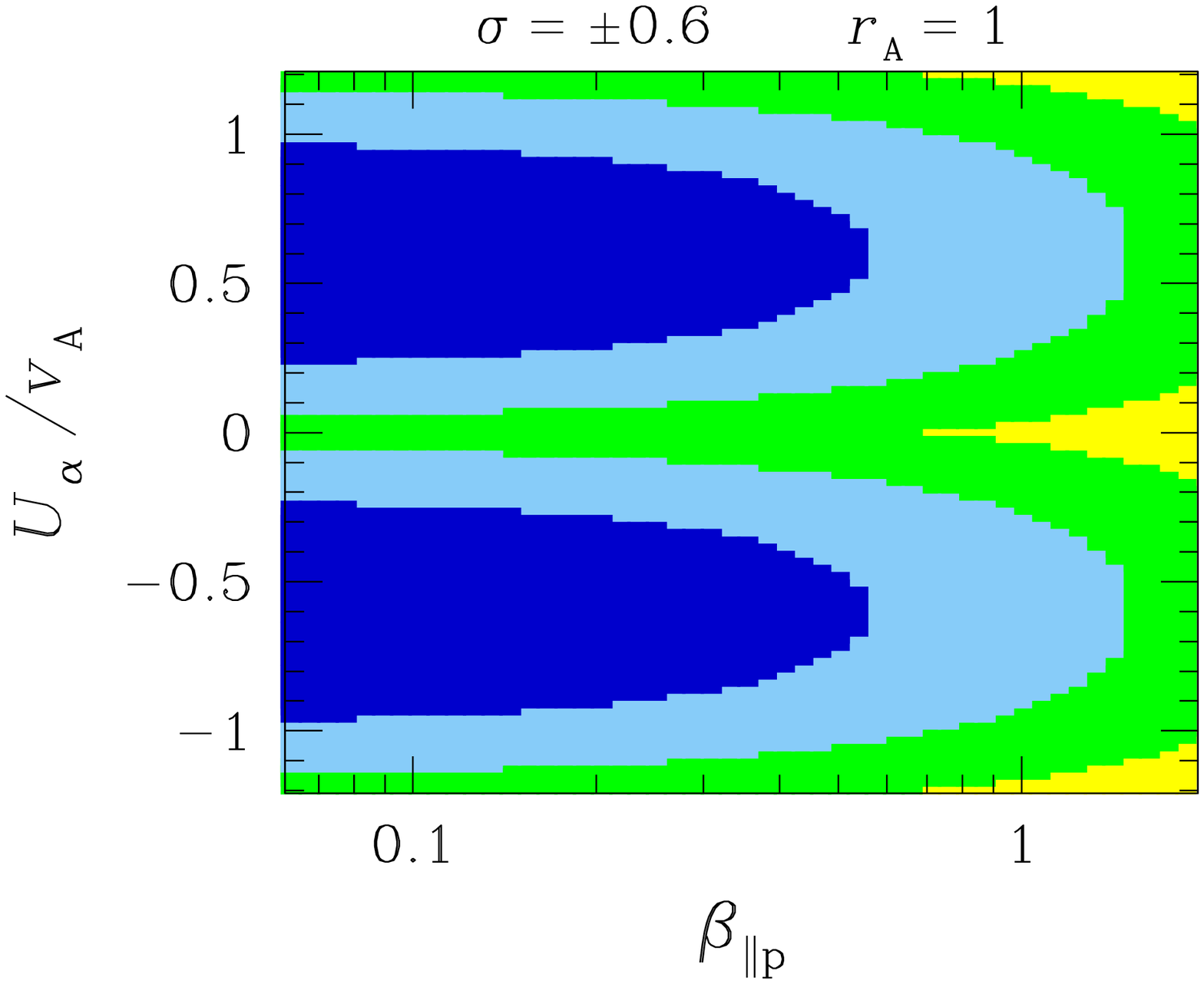}
\hspace{-4.5cm} 
\includegraphics[width=5.5cm]{color_bar.eps}
}
\vspace{-1cm} 
\centerline{
\includegraphics[width=5.5cm]{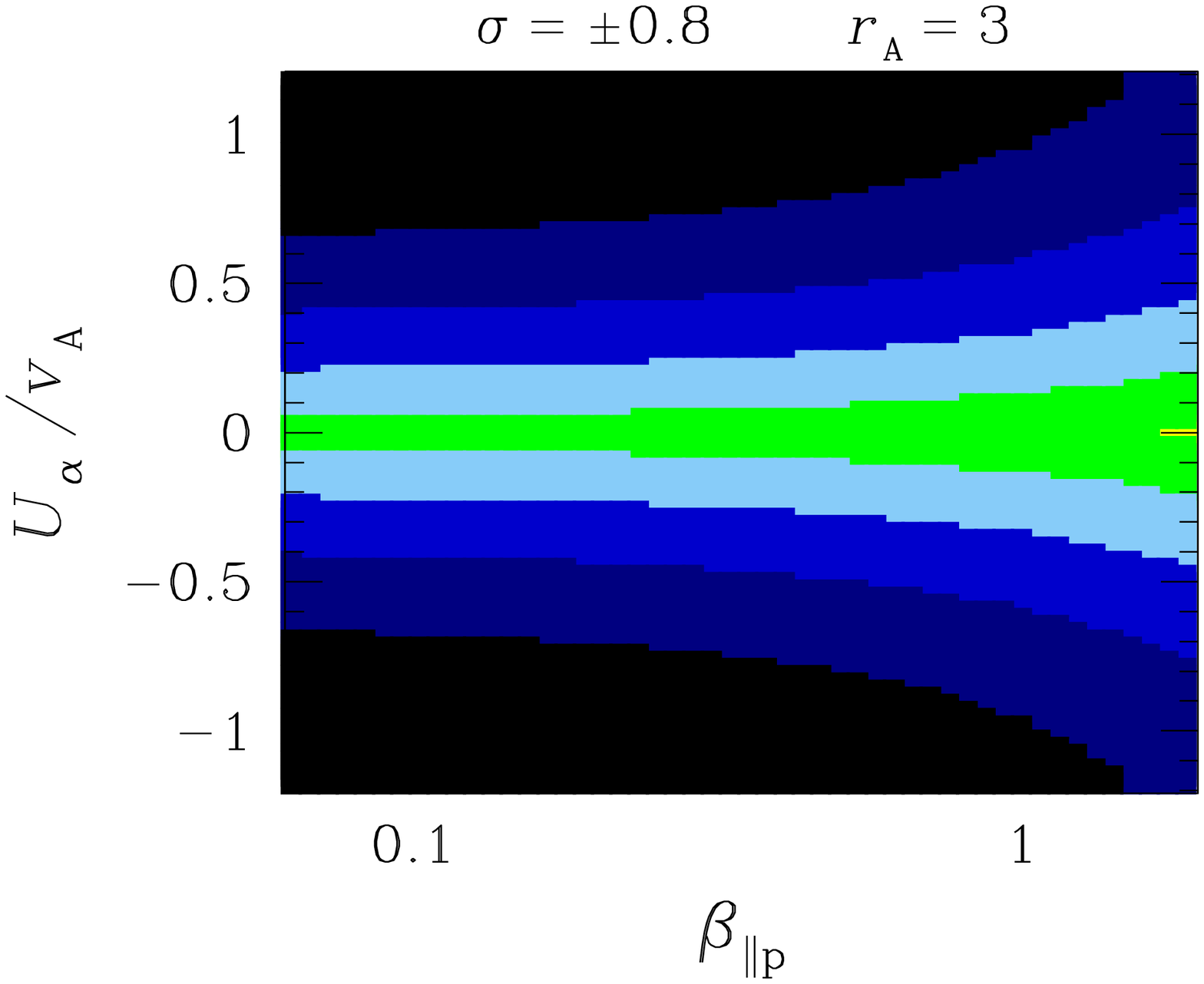}
\includegraphics[width=5.5cm]{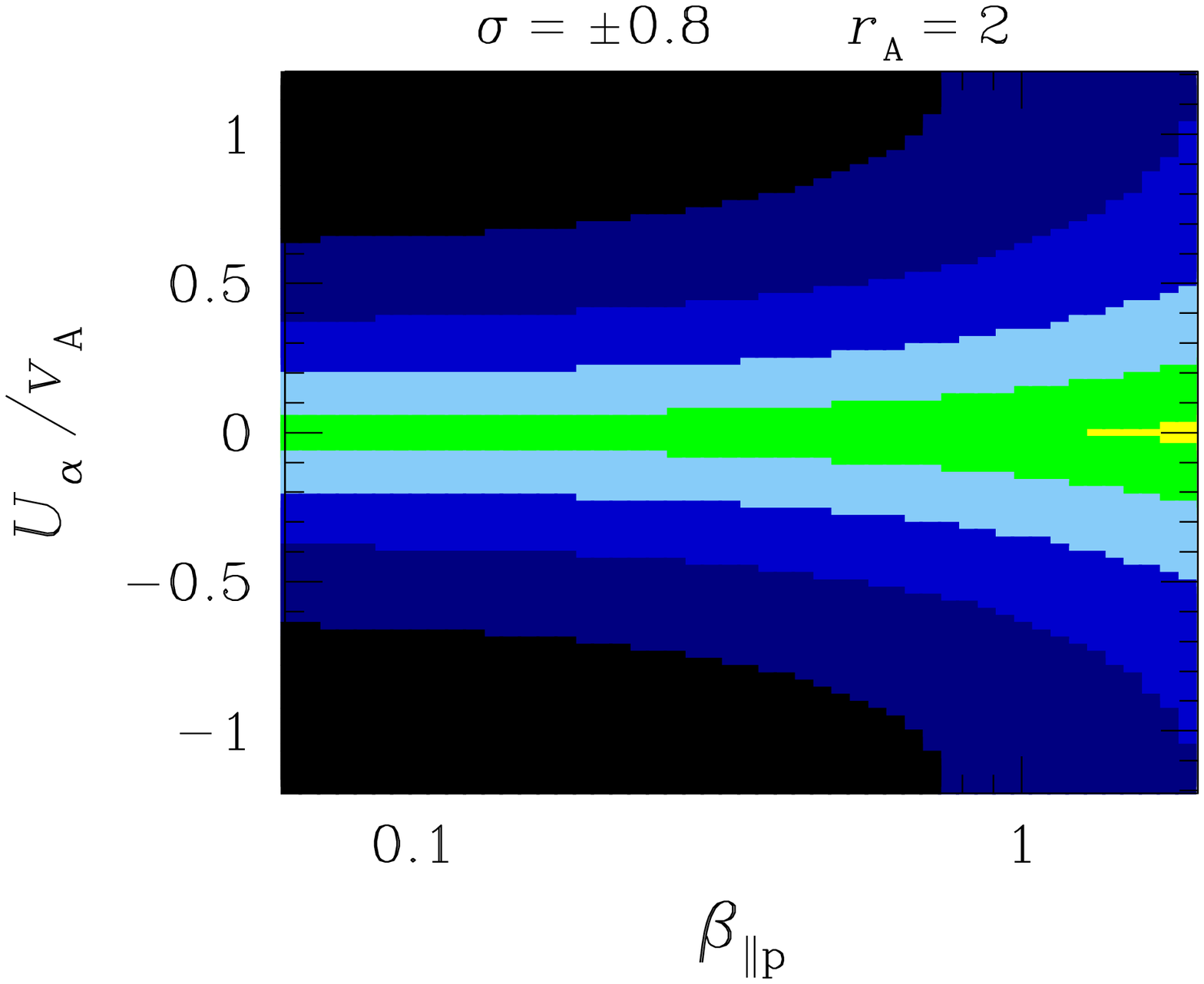}
\includegraphics[width=5.5cm]{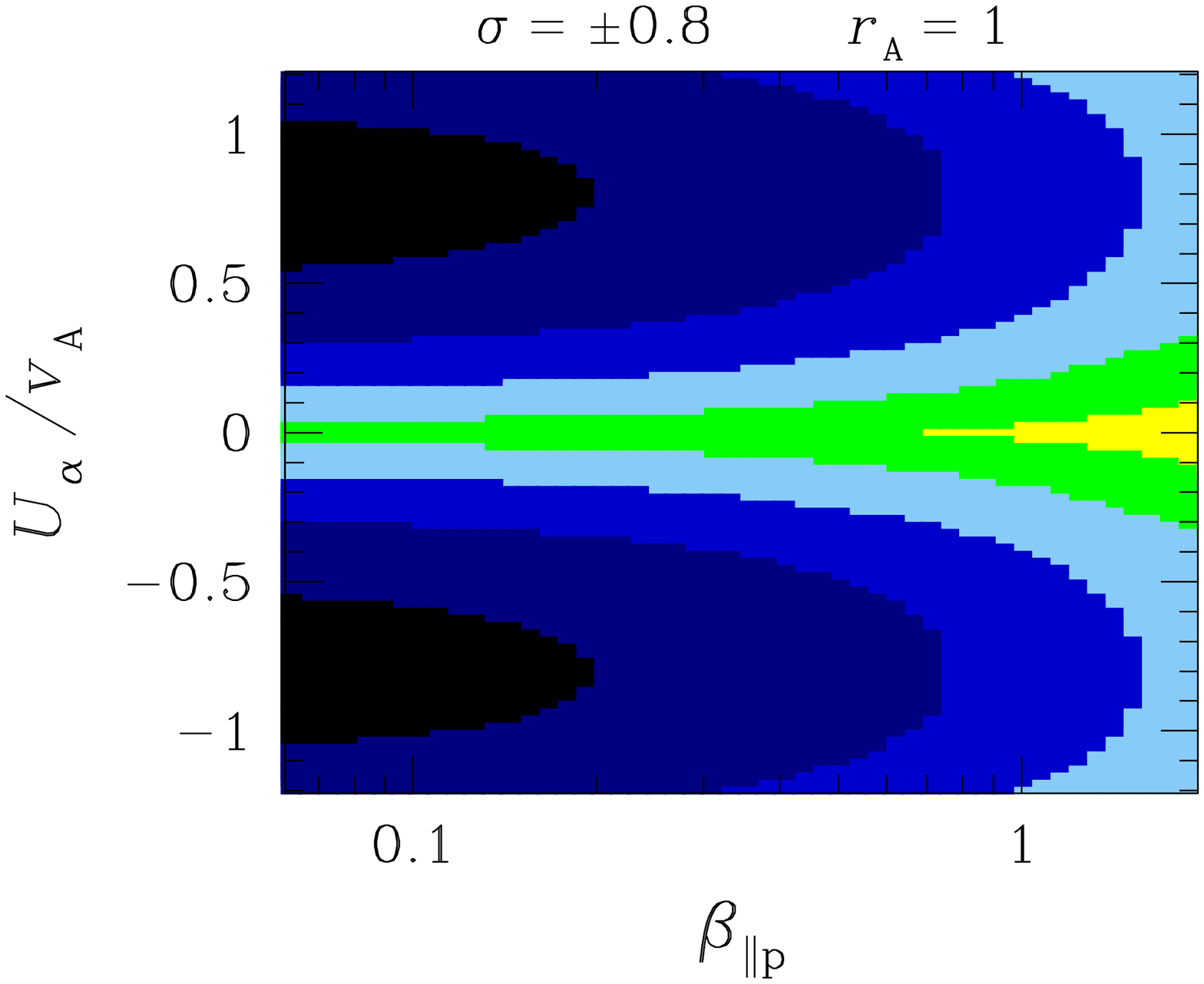}
\hspace{-4.5cm} 
\includegraphics[width=5.5cm]{color_bar.eps}
}
\vspace{-1.2cm} 
\caption{Color-scale plots of $T_{\perp \alpha}/T_{\perp \rm
    p}$ for different choices of the gyroscale fractional
  cross helicity~$\sigma$ and gyroscale Alfv\'en ratio~$r_{\rm
    A}$. For all figures, $a=0.25$ and $T_{\parallel
    \alpha}/T_{\parallel \rm p} = 5.2$. These results can be compared
  directly to the {\em Wind} measurements in Figure~2 of
  \cite{kasper13}.
\label{fig:Kasper_Talpha} }
\end{figure*}

In Figure~\ref{fig:Kasper_Talpha}, we plot $T_{\perp \alpha}/T_{\perp
  \rm p}$ as a function of $U_\alpha/v_{\rm A}$ and $\beta_{\parallel
  \rm p}$ for nine different combinations of the parameters $r_{\rm
  A}$ and~$\sigma$. For all panels of this figure, we set
$T_{\parallel \alpha}/T_{||\rm p} = 5.2$, as in
Equation~(\ref{eq:Tparratio}), and~$a=0.25$.  The portion of each plot
with $U_\alpha <0 $ is obtained by reflecting the upper half of the
plot through the line $U_\alpha = 0$. The lower half of each plot
represents cases in which $\bm{B}_0$ is directed towards the Sun, so
that $U_\alpha$ and $\sigma$ are both negative. Thus, in the
upper-right panel of Figure~\ref{fig:Kasper_Talpha} with the
label~$\sigma = \pm 0.4$, the value $\sigma = +0.4$ corresponds to the
upper half of the plot in which $U_\alpha > 0$, and the value $\sigma
= -0.4$ corresponds to the lower half of the plot in which~$U_\alpha <
0$. These plots can be compared to K13's Figure~2, which
plots, in the same coordinate plane, the average value of~$T_{\perp
  \alpha}/T_{\perp \rm p}$ in measurements of weakly collisional
solar-wind streams from the {\em Wind} spacecraft.  However, because
of Equation~(\ref{eq:betacond}), we have limited our plots to
$\beta_{\parallel \rm p} \leq 2$, whereas K13's Figure~2 
includes larger~$\beta_{\parallel \rm p}$ values.

All of the panels in Figure~\ref{fig:Kasper_Talpha} share three
features with the {\em Wind} data plotted in K13's Figure~2.
First, $ 6 \lesssim T_{\perp \alpha}/T_{\perp \rm p}
\lesssim 7$ within a band of small-$U_{\alpha}/v_{\rm A}$ values when
$\beta_{\parallel \rm p} \lesssim 1$. Second, $T_{\perp
  \alpha}/T_{\perp \rm p}$ decreases with increasing $U_\alpha/v_{\rm
  A}$ at small-$\beta_{\parallel \rm p}$, because the electric field
seen by the alpha particles is reduced when the alpha-particles drift
in the same direction that the majority of the RMHD fluctuations
propagate. Third, this decrease is less pronounced at
$\beta_{\parallel \rm p} \sim 1$, in part because the increase in
$w_{\parallel \alpha}/v_{\rm A}$ means that the averaging
over~$v_\parallel$ in Equation~(\ref{eq:Tscaling2}) increasingly
smoothes out the vertical variations in each panel, and in part
because a larger value of~$w_{\parallel \alpha}/v_{\rm A}$ means that
more alpha particles satisfy $v_\parallel < 0$ or $v_\parallel >
v_{\rm A}$, either of which conditions enhances the electric field in
the $v_\parallel$~frame.  On the other hand, several of the panels in
Figure~\ref{fig:Kasper_Talpha} exhibit enhanced $T_{\perp
  \alpha}/T_{\perp \rm p}$ values at $U_\alpha \simeq v_{\rm A}$,
which are not seen in the~{\em Wind} data. This discrepancy becomes
increasingly pronounced as $|\sigma|$ and/or $r_{\rm A}$ decrease.
The ability of our model to explain the {\em Wind} data thus depends
on the values of $\sigma$ and $r_{\rm A}$ in the solar wind,
as we discuss further in Section~\ref{sec:conclusion}.

\begin{figure*}[t]
\centerline{
\includegraphics[width=5.5cm]{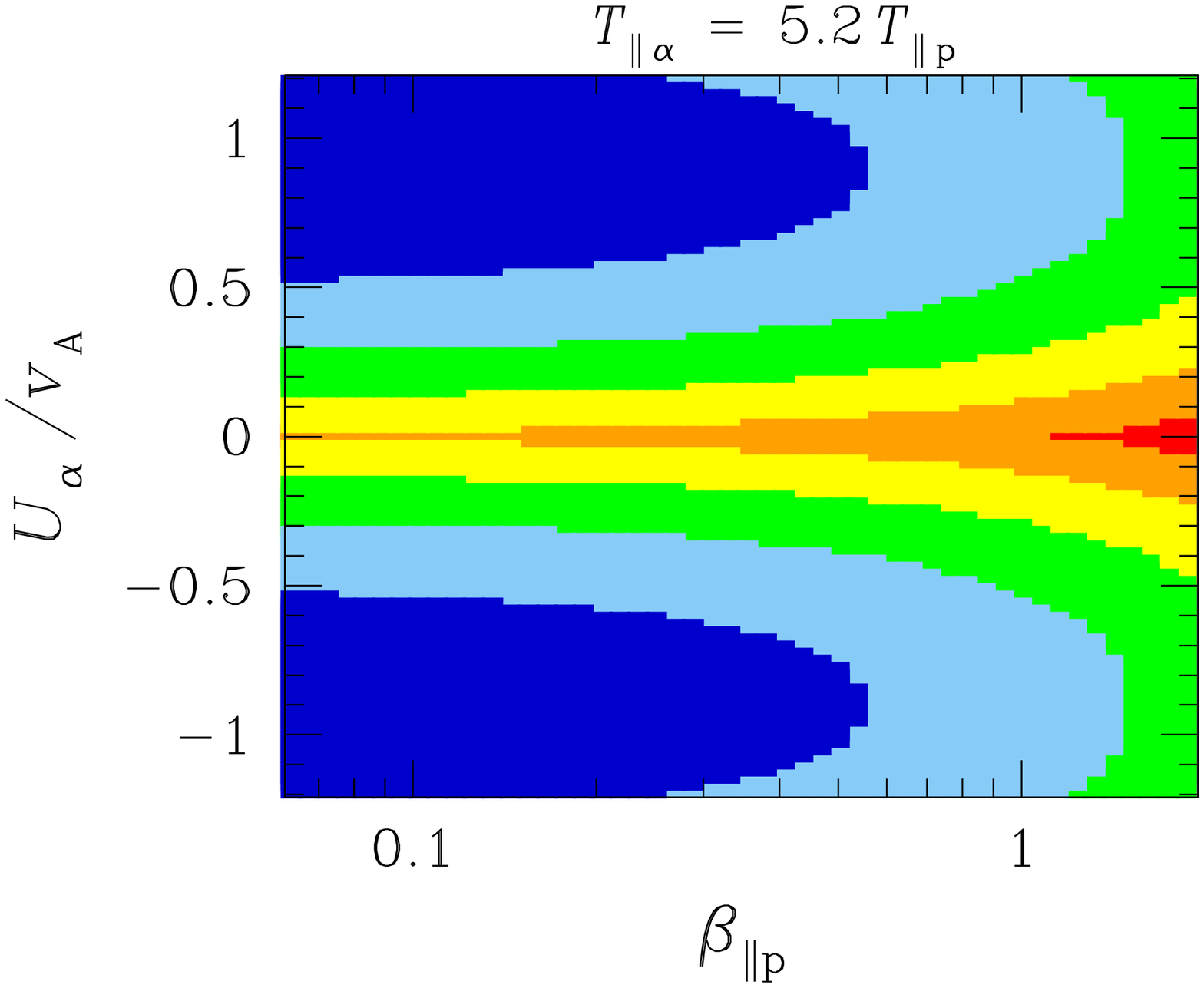}
\hspace{0.5cm} 
\includegraphics[width=5.5cm]{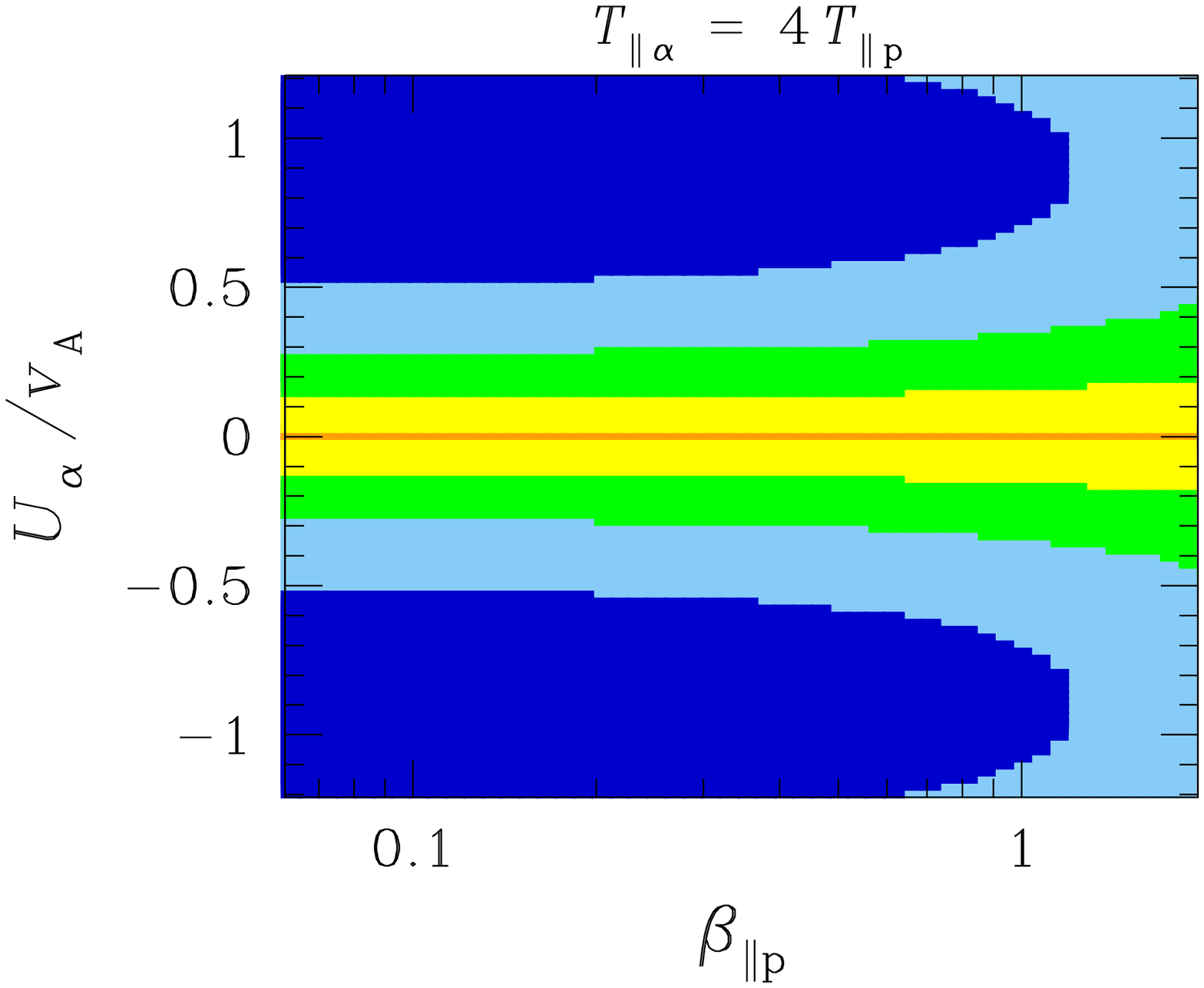}
\hspace{-4cm} 
\includegraphics[width=5.5cm]{color_bar.eps}
}
\vspace{-1cm} 
\caption{Color-scale plots of $T_{\perp \alpha}/T_{\perp \rm p}$ as a
  function of $\beta_{\parallel \rm p}$ and the average alpha-particle
  velocity~$U_\alpha$.  In both panels, the scaling exponent~$a$
  is~$0.35$, and the gyroscale Alfv\'en ratio~$r_{\rm A}$ is~2. The
  gyroscale fractional cross helicity $\sigma$ is $0.6$ in the upper
  half planes in which~$U_\alpha>0$ and $-0.6$ in the lower half
  planes in which~$U_\alpha < 0$. The parallel temperature
  ratio~$T_{\parallel \alpha}/T_{\parallel \rm p}$ is 5.2 in the left
  panel and 4.0 in the right panel.  The color bar in this figure is
  the same as in Figure~\ref{fig:Kasper_Talpha} to facilitate
  comparison of these figures, but the two darkest colors
  are not used in either panel of Figure~\ref{fig:Kasper_Talpha2}.
\label{fig:Kasper_Talpha2} }
\end{figure*}

In Figure~\ref{fig:Kasper_Talpha2} we illustrate how the color-scale
plots of Figure~\ref{fig:Kasper_Talpha} change when we use a larger
value of~$a$ or a smaller value of~$T_{\parallel \alpha}/T_{\parallel
  \rm p}$. The
left panel of Figure~\ref{fig:Kasper_Talpha2} is the same as the
middle plot in Figure~\ref{fig:Kasper_Talpha}, except that~$a$ has
been increased from~$0.25$ to~0.35.  This increase in~$a$
increases~$T_{\perp \alpha}/T_{\perp \rm p}$ at all locations in this
plot because of the change in the factor of~$(A/Z)^{2a\eta}$ on the
right-hand side of Equation~(\ref{eq:Tratio}).  The right panel of
Figure~\ref{fig:Kasper_Talpha2} is the same as the left panel of
Figure~\ref{fig:Kasper_Talpha2}, except that $T_{\parallel
  \alpha}/T_{\parallel \rm p}$ has been decreased from~5.2 to~4. The
primary effect of decreasing $T_{\parallel \alpha}/T_{\parallel \rm
  p}$ is to shift the plot towards the right.  This is because the
averaging over~$v_\parallel$ that occurs in Equation~(\ref{eq:Tratio}) 
smoothes out the vertical variations
in~$T_{\perp \alpha}/T_{\perp \rm p}$ in these plots when
$w_{\parallel \alpha} \gtrsim v_{\rm A}$, and the condition
$w_{\parallel \alpha} = v_{\rm A}$ is satisfied at
larger~$\beta_{\parallel \rm p}$ when $T_{\parallel
  \alpha}/T_{\parallel \rm p}$ is smaller.

\begin{figure}[b]
\vspace{0.5cm} 
\centerline{
\includegraphics[width=7cm]{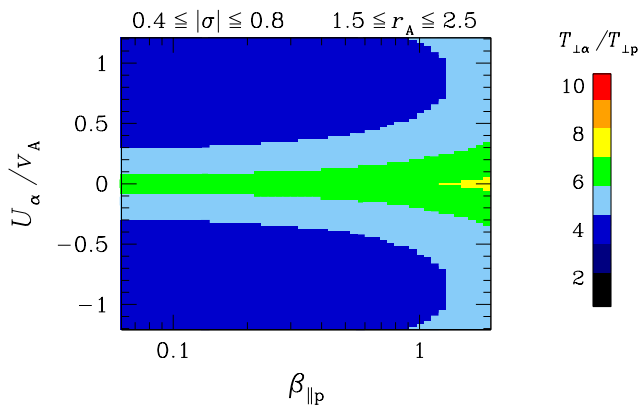}
\hspace{-5.5cm} 
\includegraphics[width=7cm]{color_bar.eps}
}
\vspace{-1.5cm} 
\caption{Color-scale plot of $T_{\perp \alpha}/T_{\perp \rm p}$ as a
  function of $\beta_{\parallel \rm p}$ and $U_{\alpha}$ averaging
  over uniform distributions of $|\sigma|$ and $r_{\rm A}$, which vary
  throughout the ranges $0.4 \leq |\sigma| \leq 0.8$ and $1.5 \leq
  r_{\rm A} \leq 2.5$.  The parallel temperature ratio~$T_{\parallel
    \alpha}/T_{\parallel \rm p}$ is assumed to be 5.2, and $a=0.25$.
  The color bar is the same as in Figures~\ref{fig:Kasper_Talpha}
  and~\ref{fig:Kasper_Talpha2} to facilitate comparison with these
  figures, but several colors that appear in this color bar
  are not needed for this plot.
\label{fig:Kasper_Talpha3} }
\end{figure}

In Figure~\ref{fig:Kasper_Talpha3}, we average $T_{\perp
  \alpha}/T_{\perp \rm p}$ over uniform distributions of $|\sigma|$
and~$r_{\rm A}$. In particular, we calculate $T_{\perp
  \alpha}/T_{\perp \rm p}$ for each of ten values of $|\sigma|$ evenly
spaced between 0.4 and~0.8 (inclusive) and each of ten values of
$r_{\rm A}$ evenly spaced between 1.5 and 2.5 (inclusive), and then we
average the resulting 100 values of $T_{\perp \alpha}/T_{\perp \rm
  p}$. For this figure, we set $a=0.25$ and $T_{\parallel \alpha} =
5.2 T_{\parallel \rm p}$.  Of all the plots in
Figures~\ref{fig:Kasper_Talpha} through~\ref{fig:Kasper_Talpha3}, 
Figure~\ref{fig:Kasper_Talpha3} is the most similar to
K13's Figure~2.

We note that K13 found enhanced values of $T_{\perp \rm
  p}/T_{\parallel \rm p}$ in the upper-left and lower-left corners of
the plane shown in each of the plots of
Figures~\ref{fig:Kasper_Talpha} through~\ref{fig:Kasper_Talpha3}. They
interpreted this result within the context of a
resonant-cyclotron-heating model as evidence that when $T_{\perp
  \alpha}/T_{\perp \rm p}$ is comparatively small (i.e., $\sim 4$
rather than~$\sim 7$), the alpha-particles drain less power from the
A/IC waves, enabling the A/IC waves to cause enhanced perpendicular
proton heating. It is possible that a similar interpretation can be
applied within the context of a stochastic-heating model. In this
case, weaker alpha-particle heating would drain less power from the
turbulent cascade, enabling more energy to cascade to the
proton-gyroradius scale and enhancing the stochastic heating of
protons. However, a direct prediction of the magnitude of this effect
is beyond the scope of this paper.

\vspace{0.2cm} 
\section{Discussion and Conclusion}
\label{sec:conclusion} 
\vspace{0.2cm} 

In this paper, we extend Chandran et al's (2010) theoretical treatment
of stochastic heating to account for the motion of ions along the
background magnetic field. Using this more general theory, we derive an
analytic expression for the ion-to-proton perpendicular temperature
ratio~$T_{\perp \rm i}/T_{\perp \rm p}$ in the solar wind under the
assumption that stochastic heating is the dominant perpendicular
heating mechanism for all ions.  This expression determines the
dependence of $T_{\perp \rm i}/T_{\perp \rm p}$ on the average drift
velocity of the ions of species~i relative to the protons, the
parallel thermal speeds of both species, the gyroscale fractional
cross helicity~$\sigma$, the gyroscale Alfv\'en ratio~$r_{\rm A}$, and
the scaling exponent~$a$ defined in Equation~(\ref{eq:Wscal}), which
characterizes the steepness of the turbulent power spectra at
wavenumbers~$\sim \rho_{\rm p}^{-1}$.  Our calculation is restricted
to values of~$\beta_{\parallel \rm p}$ that are~$\lesssim 1$ (see the
discussion preceding Equation~(\ref{eq:betacond})).

When applied to alpha particles, our results reproduce three features
of the alpha-to-proton perpendicular temperature ratios measured by
the {\em Wind} spacecraft in weakly collisional solar-wind
streams~\citep{kasper13}, at least for certain values of the model
parameters.  First, if we set $a=0.25$ (the value that would arise if
the electric-field and magnetic-field power spectra were~$\propto
k_\perp^{-3/2}$ at $k_\perp \rho_{\rm p} \sim 1$), then we find that
$6 \lesssim T_{\perp \alpha}/T_{\perp \rm p} \lesssim 7$ at small
$U_{\alpha}/v_{\rm A}$ when $\beta_{\parallel \rm p} \lesssim
1$. Second, when $\beta_{\parallel \rm p}$ is small, $T_{\perp
  \alpha}/T_{\perp \rm p} $ decreases markedly as $U_{\alpha}/v_{\rm
  A}$ increases from 0 to~1, provided $r_{\rm A}$ and $\sigma$ are not
too small. Third, this decrease becomes less pronounced as
$\beta_{\parallel \rm p}$ increases to values~$\sim 1$.  On the other
hand, if $|\sigma|$ and/or~$r_{\rm A}$ are reduced sufficiently, then
our model produces enhanced values ($\gtrsim 5$) of $T_{\perp
  \alpha}/T_{\perp \rm p}$ at $|U_{\alpha}| \sim v_{\rm A}$ that are
not seen in the data. The ability of our model to explain the data
thus depends on the values of $\sigma$ and $r_{\rm A}$ in the solar
wind.  Of all the $T_{\perp \alpha}/T_{\perp \rm p}$ plots shown in
this paper, the one that most closely resembles K13's Figure~2 is our
Figure~\ref{fig:Kasper_Talpha3}. In this panel, $T_{\perp
  \alpha}/T_{\perp \rm p}$ is averaged over uniform distributions
of~$\sigma$ and~$r_{\rm A}$ in which $0.4 \leq |\sigma| \leq 0.8$ and
$1.5 \leq r_{\rm A} \leq 2.5$.

Observationally, the distribution of~$\sigma$ and $r_{\rm A}$ values
in the solar wind is not clear. The difficulty is that these
quantities depend upon the $\bm{E}$ and $\bm{B}$ fluctuations at
scales~$\sim \rho_{\rm p}$, and at these scales the $\bm{E}$
measurements are noisy. That being said, there are some indications
that the values of $\sigma$ and $r_{\rm A}$ in the solar wind are in
the range of values in which our model compares well with the data.
For example, using {\em Wind} data, \cite{podesta10} found that the
fractional cross helicity is relatively constant within the inertial
range, implying that the gyroscale fractional cross helicity~$\sigma$
is typically similar to the fractional cross helicity at large
lengthscales corresponding, e.g., to $\sim 1$-hour timescales in
spacecraft measurements. Fractional cross helicities at $\sim 1$-hour
timescales in the range of 0.4 to~0.8 are common in the solar wind at
$r =1 \mbox{ AU}$ \citep[e.g.,][]{roberts87,bavassano00,chen13a}. Regarding
the gyroscale Alfv\'en ratio, measurements from the {\em Cluster}
spacecraft suggest that $r_{\rm A}$ is slightly greater than~1 at
lengthscales~$\sim \rho_{\rm p}$ and that $r_{\rm A}$ increases to
larger values at even smaller lengthscales~\citep{bale05,salem12}, a
finding that is similar to results from gyrokinetic simulations of
low-frequency plasma turbulence~\citep{howes08b}. However, the results
of these studies are not fully conclusive, because of the noise in the
electric-field data at scales~$\rho_{\rm p}$ and because gyrokinetic
simulations have not yet resolved a large enough range of lengthscales
bracketing $\rho_{\rm p}$ to guarantee that the physics at scales
$\sim \rho_p$ is insensitive to both the large-scale driving and
grid-scale dissipation.  Future observational and/or numerical studies
to characterize more precisely the electric-field fluctuations at
lengthscales~$\sim \rho_{\rm p}$ would lead to more rigorous tests of
the stochastic-heating model.

Further work is also needed to improve our theoretical treatment of
stochastic heating. For example, to simplify the calculation, we have
assumed that $\sigma$ and $r_{\rm A}$ are independent of scale at
scales~$\sim \rho_{\rm p}$. In the solar wind, however, $r_{\rm A}$
and possibly~$\sigma$ vary with scale.  As in \cite{chandran10a}, we
have assumed that stochastic heating is dominated by fluctuations at
scales comparable to an ion's gyroradius. In reality, however, an ion
is likely heated by fluctuations with a range of lengthscales.  In
addition, we have neglected the possible effects of
temperature-anisotropy instabilities. Such instabilities place 
upper and lower limits on $T_\perp/T_\parallel$ for both protons and alpha
particles, which could affect the value of $T_{\perp \alpha}/T_{\perp
  \rm p}$. These instabilities become increasingly important as
$\beta_{\parallel \rm p}$ increases and should be included in future
models.

Finally, although we have focused on alpha particles and protons, we
note that our analysis also applies to minor ions.  Observations
suggest that minor ions at $r= 1 \mbox{ AU}$ have $w_{\parallel \rm
  i}$ values that are similar to $w_{\parallel
  \alpha}$~\citep{bochsler07}.  Figure~\ref{fig:Kasper_Talpha}
therefore approximately describes minor ions as well as alpha
particles, provided one adjusts the color scale to account for the
change in the normalization of $T_{\perp \rm i}/T_{\perp \rm p}$ for
different ion species due to the different values of the factor
$A(A/Z)^{2a\eta} $ on the right-hand side of
Equation~(\ref{eq:Tratio}).  Measurements of $T_{\perp \rm i}/T_{\perp
  \rm p}$ for ions other than alpha particles, and how $T_{\perp \rm
  i}/T_{\perp \rm p}$ depends upon $w_{\parallel \rm i}$ and $U_{\rm
  i}$, would thus lead to further tests of the stochastic heating
model.

\acknowledgements We thank A. Schekochihin for helpful discussions and
the referee for valuable suggestions.  This work was supported by
grant NNX11AJ37G from NASA's Heliophysics Theory Program, NASA grant
NNN06AA01C to the Solar Probe Plus FIELDS Experiment, NASA grant
NNX13AF97G, NSF grant AGS-0851005, NSF/DOE grant AGS-1003451, and DOE
grant DE-FG02-07-ER46372.  B.~Chandran was supported by a Visiting
Research Fellowship from Merton College, University of Oxford.
E.~Quataert was supported by a Simons Investigator award from the
Simons Foundation, the David and Lucile Packard Foundation, and the
Thomas Alison Schneider Chair in Physics at UC Berkeley.

\bibliography{articles}

\end{document}